\documentclass[]{aa} 

\usepackage{graphicx}
\usepackage{txfonts}
\usepackage{amsmath}	
\usepackage{amssymb}	
\usepackage{booktabs}   
\usepackage{multirow}   
\usepackage{rotating}   %
\usepackage{natbib}
\usepackage{color, soul} 
\soulregister\cite7
\soulregister\citet7
\soulregister\citep7
\soulregister\ref7
\usepackage{xcolor} 
\usepackage{xspace}
\usepackage{subcaption}     
\usepackage[stable]{footmisc}
\newcommand{\flagfont}[1]{\fontfamily{cmss}{\selectfont #1}}  
\newcommand{\pyfont}{\fontfamily{cmtt}\selectfont}

\newcommand{\edit}[1]{#1}
\newcommand{\final}[1]{#1}

\begin{document}

\title{Towards characterising rocky worlds: \\ 
	Trends in chemical make-ups of M dwarfs versus GK dwarfs}
\titlerunning{Trends in chemical make-ups of M dwarfs versus GK dwarfs} 
\authorrunning{H. S. Wang et al.}

   \author{Haiyang S. Wang\inst{1,2,3,4},
          Sascha P. Quanz\inst{1,2,5}, 
          Suvrath Mahadevan\inst{1,6,7},
          and Morgan Deal\inst{8}}

   \institute{Institute for Particle Physics and Astrophysics, ETH Z\"{u}rich, 
   	          Wolfgang-Pauli-Strasse 27, 8093 Z\"{u}rich, Switzerland \\
              \email{haiwang@ethz.ch; sascha.quanz@phys.ethz.ch}
         \and
         National Center of Competence in Research PlanetS, 
         Gesellschaftsstrasse 6, 3012 Bern, Switzerland
         \and
         Institute of Geochemistry and Petrology, ETH Z\"{u}rich, 
         Clausiusstrasse 25, 8092 Z\"urich, Switzerland
         \and
         Centre for Star and Planet Formation, Globe Institute, University of Copenhagen, \O ster Voldgade 5-7, 1350 Copenhagen, Denmark \\ \email{haiyang.wang@sund.ku.dk}
         \and ETH Zurich, Department of Earth Sciences, Sonneggstrasse 5, 8092 Zurich, Switzerland
         \and
         Department of Astronomy \& Astrophysics, 525 Davey Laboratory, The Pennsylvania State University, University Park, PA 16802, USA
         \and
         Center for Exoplanets and Habitable Worlds, 525 Davey Laboratory, The Pennsylvania State University, University Park, PA 16802, USA
         \and
         LUPM, Universit\'e de Montpellier, CNRS, Place Eug\`ene Bataillon, 34095 Montpellier, France
     }

   \date{Received xxx; accepted xxx}

  \abstract
   {\edit{Elemental abundances of Sun-like stars have been shown to be crucial for understanding the detailed properties of planets surrounding them. However, accurately measuring elemental abundances of M stars, the most abundant class of stars in the Solar neighbourhood, is challenging due to their faintness and pervasive molecular features in optical photospheric spectra. As a result, elemental abundances of Sun-like stars have been proposed to constrain those of M stars, particularly by scaling [X/H] given measured [Fe/H].}} 
   {\edit{This work aims to test the robustness of this convenient practice based on two selected sets of M- and GK-dwarf stellar abundances and a set of rigorous statistical methods.}}
   {We compile the elemental abundances of a sample of up to 43 M dwarfs for 10 major rock-forming elements (Fe, C, O, Mg, Si, Al, Ca, Na, Ni, and Ti) \edit{from high-resolution near-infrared stellar surveys including APOGEE, CARMENES and Subaru}. We carry out bootstrap-based linear regressions on the selected sample of M dwarfs to \edit{constrain the} statistical trends of [X/H] vs. [Fe/H] and then compare them with those of GK dwarfs (sampled from the GALAH database). \edit{We then apply a 2-sample, multivariate Mahalanobis Distance test to assess the significance of the differences of the [X/H]--[Fe/H] trends for individual elemental pairs between M and GK dwarfs.}}
   {We find that the null hypothesis -- i.e. no significant difference in the chemical trends of [X/H] vs. [Fe/H] between M and GK dwarfs -- is \final{strongly} rejected for all elements \final{except for Si, for which the rejection is marginal ($p$-value close to 0.05), and Na and Ni, for which the results are inconclusive}. This finding suggests that assuming no difference may result in biased results and thus inaccurate constraints on characterising rocky planets around M dwarfs by scaling the (unmeasured) chemical abundances of planet-hosting M dwarfs from the chemical trends of [X/H]--[Fe/H] determined by GK dwarfs.} 
   {It is therefore crucial for both the stellar and exoplanet communities to be aware of these observed differences. To better understand these differences, we advocate for dedicated modelling techniques for M dwarf atmospheres and an increasing set of benchmark, homogeneous abundances analyses. Intermediately, our statistically constrained trends of [X/H]--[Fe/H] \edit{for M dwarfs provide a new constraint on estimating M-dwarf elemental abundances given measured [Fe/H] and further on characterising the detailed properties of M dwarf-hosted rocky worlds.}} 
   
   
\maketitle
%

\section{Introduction}
\edit{M stars are the most abundant type of stars in the Solar neighbourhood \citep{Reyle2021, Golovin2023} and have been among the primary stellar targets to find rocky planets (such as the TRAPPIST-1 planets; \citealt{Gillon2017}). These planets are most suitable for detailed characterisation with current observational techniques, such as, transiting photometry, radial velocity, and transmission spectroscopy \citep{Sebastian2021, Ribas2023, Lustig-Yaeger2023a}. However, due to the faintness of such cool stars and, more importantly, the pervasive molecular features present in their photospheric spectra particularly at optical wavelengths, it is a known challenge to accurately measure the photospheric elemental abundances of M dwarfs \citep{Reiners2018}.} 

\edit{On the other hand, the elemental} abundances of hundreds of thousands of Sun-like stars (particularly GK dwarfs) have been made available with high-quality photospheric spectroscopy \edit{\citep{Adibekyan2012, Bensby2014, Buder2021}} and shown to be crucial to constrain the properties of planets \edit{or planetary building blocks} around them \citep[e.g.][]{Hinkel2018, Putirka2019, Wang2019b, Bitsch2020, Wang2022, Wang2022b, spaargaren2023, Timmermann2023}. \edit{As a result, the broadly available} elemental abundances of Sun-like stars have been \edit{employed} to \edit{help} constrain the plausible elemental abundances of M dwarfs, \edit{based on which the characterisation of planets around M dwarfs is further made} \citep[e.g.][]{Unterborn2018}. 

\edit{Near-infrared spectra of M stars provide higher flux levels and simultaneously suppress molecular absorption features that may seriously distort the atomic lines of many elements \citep{Reiners2018}. Indeed, a series of benchmark sets of high-quality M dwarf abundances has been available}, thanks to high-resolution near-infrared spectrograph surveys, such as, APOGEE \citep{Souto2022}, CARMENES \citep{Ishikawa2020}, and Subaru/IRD \citep{Ishikawa2022}. These available M dwarf abundances permit a statistical examination of the difference and similarity of the chemical make-ups between M dwarfs and \edit{(Sun-like) GK dwarfs}. Further, any chemical trends \edit{resultant from analysing these benchmark sets of high-quality data} will improve our knowledge of M dwarf chemical compositions and, by extension, the compositions of planets around such cool stars.

\edit{The paper is organised as follows. In Sect. \ref{sec:data}, we describe our sample selections for both GK and M dwarfs and then show the selected data in the [X/H]--[Fe/H] space with a preliminary analysis of the correlation of individual elemental pairs. In Sect. \ref{sec:analysis}, we present our main analysis and results of the chemical trends of both GK and M dwarfs and the statistical tests of the consistency between these trends. In Sect. \ref{sec:discussion}, we discuss the effect of different sample selections, plausible reasons for observed differences, as well as the potential impact of such observed differences on characterising planets around M dwarfs. We conclude in Sect. \ref{sec:conclusion}. It is necessary to note that, in order to keep the manuscript concise, we present in our main figures only four out of nine pairs of elements that we analysed. The full sets of nine-panel plots for the various analyses are shown in Appendix \ref{app:full_compar}. We have also offered additional information on population classification of the samples of GK dwarfs in Appendix \ref{app:class}, the implementation details of the 2-sample multivariate Mahalanobis Distance test (with open-access codes) in Appendix \ref{app:mahala}, and three sets of supplementary analyses for supporting the main analysis in Appendices \ref{app:150pc}--\ref{app:sens_test}.} 

\section{Sample Selection}
\label{sec:data}
\subsection{Sample of GK dwarfs}\label{sec:data_GK}

\begin{figure*}[htbp!]
	\centering 
	\includegraphics[trim=0cm 0cm 0cm 0cm, scale=0.6, angle=0]{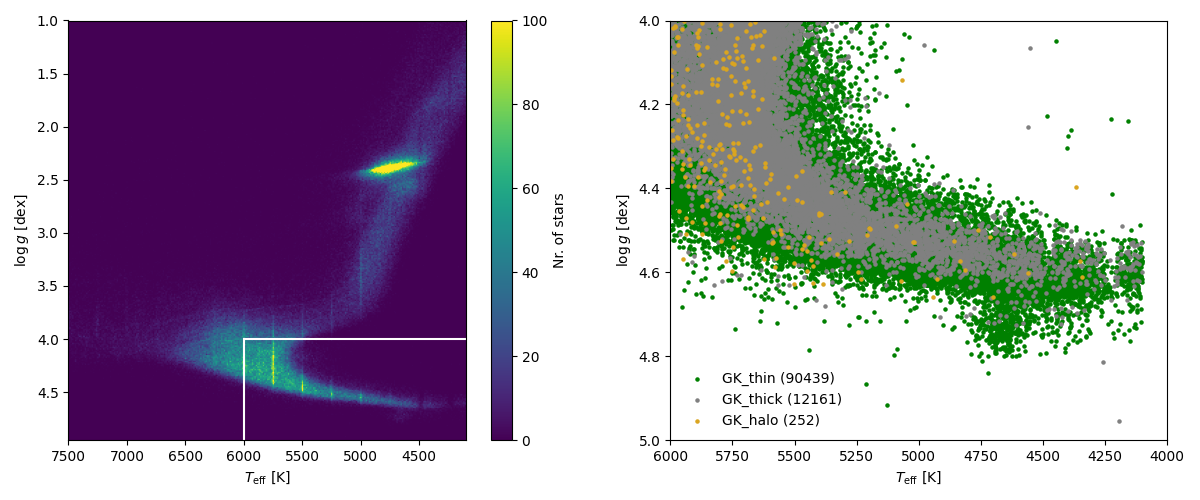} 
	\caption{Sample selection for Sun-like stars from the GALAH DR3 database \citep{Buder2021}. Left: all stars (325,518 in total) with reliable stellar parameters (flag\_sp = 0 and flag\_fe\_h = 0) and S/N > 30 (snr\_c3\_iraf > 30) on a Kiel diagram colored by the number density of stars per pixel. The GK dwarfs (\edit{102,852}) are indicated by a rectangle at the right bottom and further shown on the right panel. Right: Those GK dwarfs are further classified into three stellar populations \edit{(see Appendix \ref{app:class})}: thin-disc stars (\edit{90439}; in green), thick-disc stars (\edit{12161}; in grey), and halo stars (\edit{252}; in yellow). Only thin-disc GK dwarf stars are adopted for subsequent analysis.} 
	\label{fig:galah_data_param}
\end{figure*}

\begin{figure}[htbp!]
	\centering 
	\includegraphics[trim=0cm 0cm 0cm 0cm, scale=0.3, angle=0]{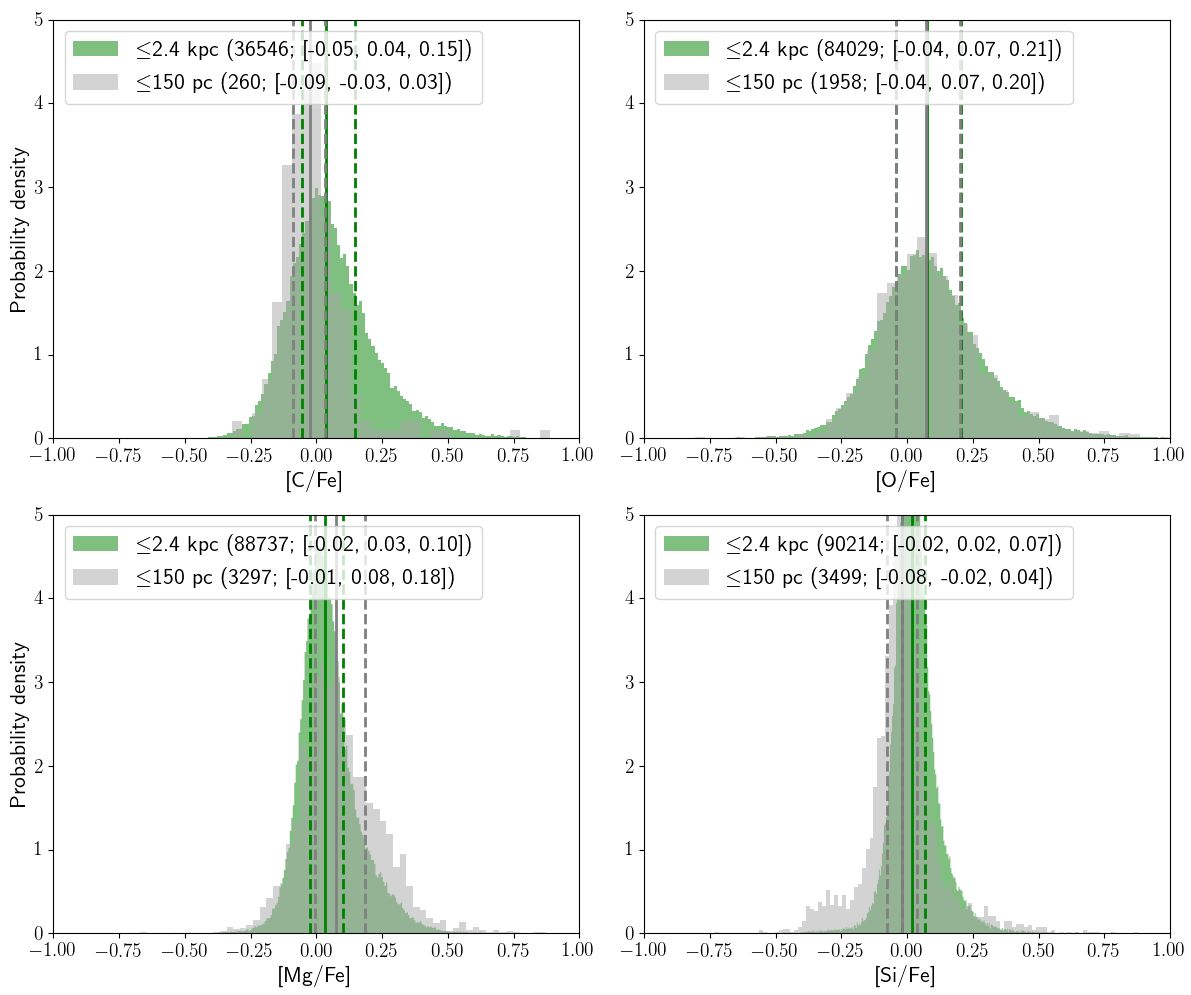} 
	\caption{\edit{Comparison of the probability densities of [X/Fe] between the full sample of thin-disc GK dwarfs (with a distance of up to 2.4 kpc from the Sun; in green) and the smaller subset of thin-disc GK dwarfs within 150 pc (in grey) of the Solar neighborhood. In the parentheses in the legend, the first value is the number of stars with valid abundances of element X, and the values within brackets correspond to [X/Fe] values at [25, 50, 75] percentiles. Such [X/Fe] values are also indicated by solid (50 percentile) and dash (32 and 68 percentiles) lines in green (for the full "$\leq 2.4$ kpc" sample) and in grey -- for the nearby "$\leq150$ pc" sample. Here we show only the cases for C, O, Mg, and Si; the full set of comparisons for all elements can be found in Fig. \ref{fig:DR3_data_hist_main}.}}
	\label{fig:DR3_data_hist_COMgSi}
\end{figure}

Our sample of Sun-like stars (G and K dwarfs)\footnote{We exclude F dwarfs from our sample selection, as they are younger and suffer stronger atomic diffusion \citep{Deal2017, Deal2018}} 
is selected from the GALAH DR3 catalogue \citep{Buder2021} based on both the stellar effective temperature ($4000 \mathrm{K} < T_{\mathrm{eff}} < 6000$ \edit{K}) and surface gravity ($\log g > 4\,\mathrm{dex}$) (the left panel of Fig. \ref{fig:galah_data_param}). The GALAH database is preferably adopted for its advantage in a homogeneous analysis of a vast amount of various stars in the Milky Way (\edit{with a distance of} up to 2.4 kpc \edit{from the Sun}). We classify the GK dwarfs (102,852 in total) 
into thin-disc stars (\edit{90,439}), thick-disc stars (\edit{12,161}), and halo stars (\edit{252}) (right panel of Fig. \ref{fig:galah_data_param}) based on probability calculations using the approach of \cite{Bensby2014} \edit{(see Appendix \ref{app:class} for details). Only thin-disc GK dwarf stars are adopted for subsequent analysis.} 

For the selection of stellar abundances ([X/H]) for 10 major rock-forming elements (C, O, Mg, Si, Al, Ca, Na, Ni, Ti, and Fe)\footnote{Sulphur is also a crucial rock-forming element, but its abundances are not available in the GALAH database, as the triplet S I lines (around 6756 $\AA$), commonly used for sulphur abundance determination, fall out of the discrete optical wavelength bands of the GALAH/HERMES spectrograph \citep{Buder2021}.}, only those with a reliable abundance flag ({\flagfont flag\_x\_fe==0}) in the GALAH DR3 catalogue are adopted. This selection yields different numbers of (thin-disc) GK dwarfs for different pairs of elements ([X/H] vs. [Fe/H]), varying from the minimum \edit{36,546} for carbon and the maximum \edit{90,326} for sodium. 

\edit{While our selected M dwarfs (see detailed description below) are within 150 pc of the Solar neighbourhood, we have decided to keep our full sample of GK dwarfs (up to 2.4 kpc) for the following two reasons: (1) if we were to confine our sample of GK dwarfs to lie within 150 pc, the available number of those stars in the GALAH database would be reduced significantly, by $\sim$ 96\%; (2) we found that the distributions of the "nearby" (150-pc) subset of GALAH (thin-disc) GK dwarfs and the full sample of GALAH (thin-disc) GK dwarfs are actually similar in the [X/Fe] space (see Fig. \ref{fig:DR3_data_hist_COMgSi} and Fig. \ref{fig:DR3_data_hist_main}), except for sodium (and, to some extent, carbon). Considering the completeness of the available GK dwarf sample and that only the GK dwarfs belonging to thin-disc populations (see Appendix \ref{app:class}) are included in our analysis, we adopt the full sample of GALAH (thin-disc) GK dwarfs for this study. For further discussion on the effect of different sample selections, see Sect. \ref{sec:discussion_sample}.} 

\subsection{Sample of M dwarfs}
\begin{figure}[htbp!]
	\centering 
	\includegraphics[trim=0cm 0cm 0cm 0cm, scale=0.3, angle=0]{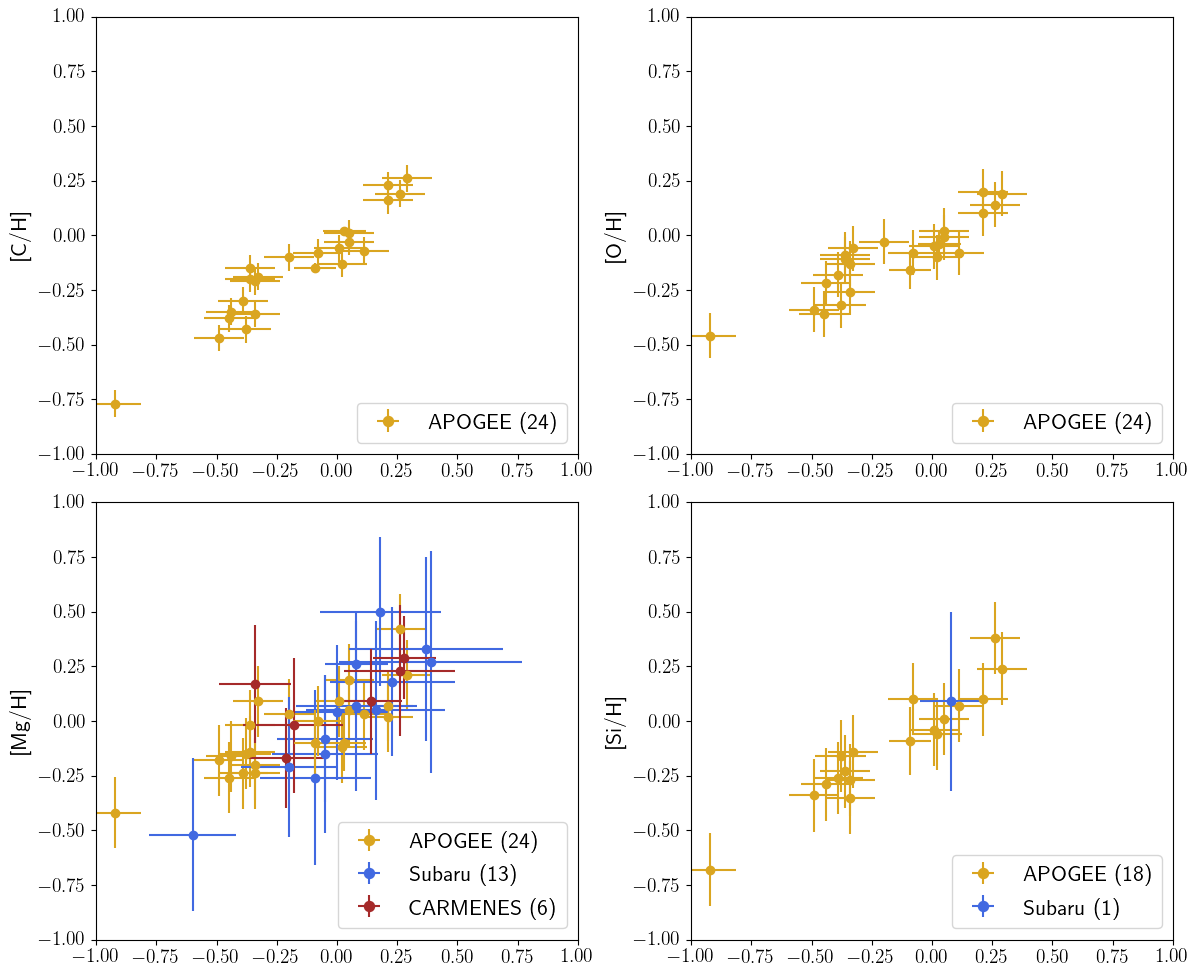} 
	\caption{\edit{Exemplary elemental abundances of the selected sample of M dwarfs (up to 43 in total) in the [X/H]--[Fe/H] space, compiled from the APOGEE \citep{Souto2017, Souto2018, Souto2022}, Subaru/IRD \citep{Ishikawa2022}, and CARMENES \citep{Ishikawa2020} surveys. The numbers of M dwarfs from these individual surveys are shown for individual elemental pairs. See Fig. \ref{fig:M_data_scatter_werr} for the full set of selected M-dwarf abundance data.}}
	\label{fig:M_data_scatter_werr_COMgSi}
\end{figure}

Our sample of M dwarfs is selected from three surveys: APOGEE \citep{Souto2017, Souto2018, Souto2022}, CARMENES \citep{Ishikawa2020}, and Subaru/IRD \citep{Ishikawa2022}. The \edit{benchmark} stellar abundance analyses with high-resolution near-infrared spectra are made for these surveys, in using the same MARCS model atmosphere \citep{Gustafsson2008} (although the spectral line lists are different and the near-infrared wavelength coverages slightly vary). The stellar abundances of the individual samples of M dwarfs in the [X/H]-[Fe/H] space are shown to be consistent between these survey results (where abundance analyses are all available; \edit{see Fig. \ref{fig:M_data_scatter_werr_COMgSi} for the compilation for C, O, Mg and Si only and Fig. \ref{fig:M_data_scatter_werr} for the full set of M-dwarf abundance data}). We therefore combine these survey results to form a selected sample of M dwarfs (up to 43 in total) for our subsequent analysis.

\subsection{Comparison of the selected samples of GK and M dwarfs}\label{sec:sample_compare}
\begin{figure}[htbp!]
	\centering 
	\includegraphics[trim=0cm 0cm 0cm 0cm, scale=0.32, angle=0]{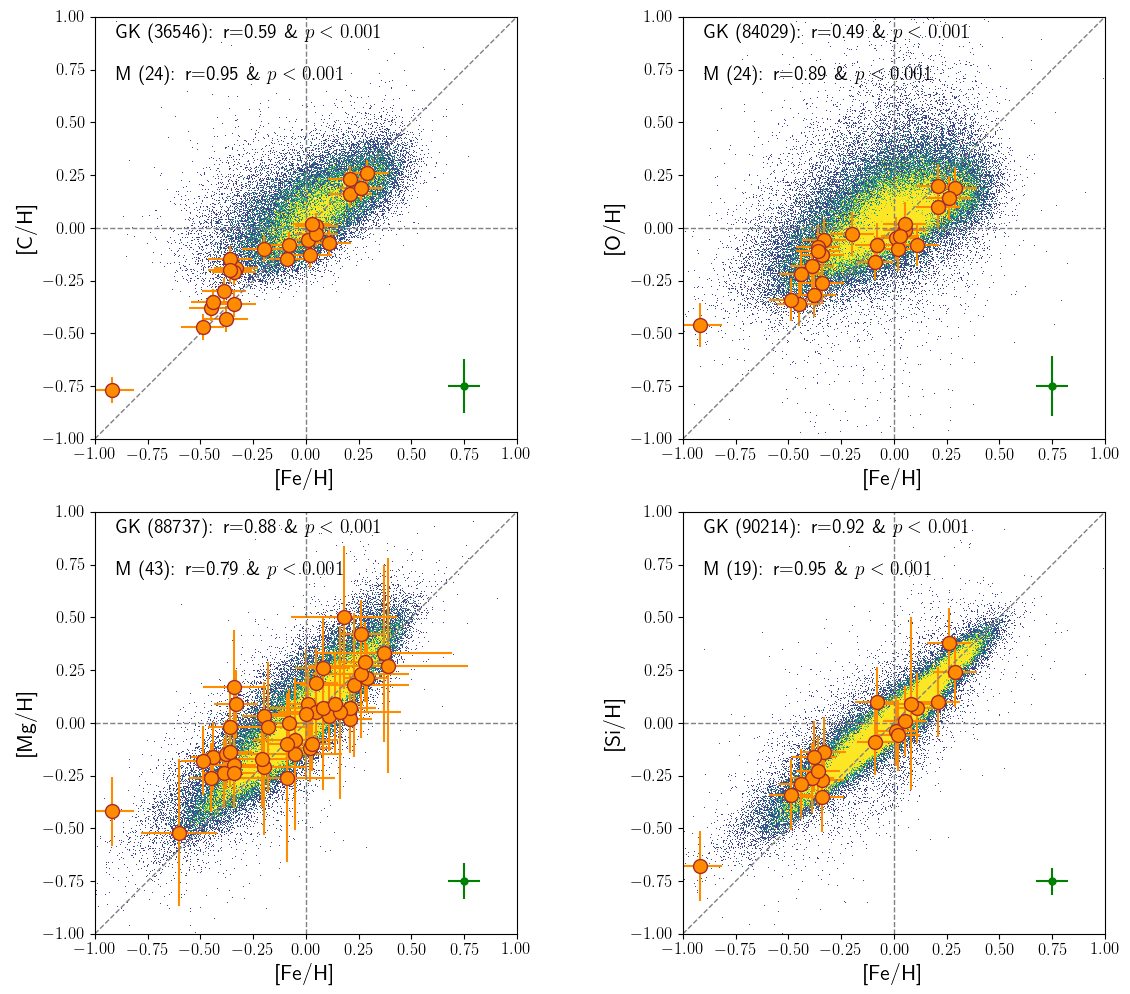} 
	\caption{\edit{Comparison of the selected samples of (thin-disc) GK dwarfs (the density plot in the background) and M dwarfs (the overlying orange dots) in the [X/H]-[Fe/H] space. The sample sizes of GK and M dwarfs are shown on each diagram, along with the corresponding test parameters (r and p-value) of Pearson's linear correlation \citep{Schober2018}. For reference, the typical error bars of the GK dwarf abundances ([X/H] vs [Fe/H]) are shown in the lower right corner of each panel, and the zero point of each panel is indicated by the horizontal and vertical dashed lines, along with the diagonal line of the panel. Here we show only the cases for C, O, Mg, and Si; the full set of comparisons for all elemental pairs can be found in Fig. \ref{fig:DR3_M_scatter_withseldata}.}}
	\label{fig:DR3_M_scatter_withseldata_COMgSi}
\end{figure}

\edit{The comparison of the selected samples of GK and M dwarfs is shown in Fig. \ref{fig:DR3_M_scatter_withseldata_COMgSi} (showing C, O, Mg and Si only, with the full set of comparisons in Fig. \ref{fig:DR3_M_scatter_withseldata}). To understand the potential correlation between [X/H] and [Fe/H] with these selected samples, we first carry out a Pearson's Linear Correlation test \citep{Schober2018}, with the test results shown in Fig. \ref{fig:DR3_M_scatter_withseldata_COMgSi} (and Fig. \ref{fig:DR3_M_scatter_withseldata}). The test results} suggest that [X/H] and [Fe/H] are strongly correlated ($r \gtrsim 0.8 $ \edit{and $p$-$value < 0.001$}) for the selected sample of M dwarfs for all elements \final{of interest} and also for the selected sample of GK dwarfs for \final{those} elements other than carbon and oxygen. \edit{For the cases of carbon and oxygen in the GK-dwarf sample}, their correlation coefficients in the [X/H]-[Fe/H] space are comparably smaller \edit{($r\mathrm{(C)} = 0.59$ and $r\mathrm{(O)} = 0.49$), but the correlations remain to be significant (with $p$-$value < 0.001$). This preliminary test guides our subsequent resampling and linear regression analyses.} 

\section{Analysis and Results}
\label{sec:analysis}

\begin{table}[htbp!]
	\caption{Final regression results (slope $\alpha$, intercept $\beta$ \edit{and coefficient of determination $R^2$}) of [X/H]-[Fe/H] trends for both M and GK dwarfs.} 
	\begin{tabular}{cccc}	
	\hline\hline
 
X & Coefs. & M dwarfs & GK dwarfs \\ 
\multirow{3}{*}{C} &$\alpha$ &$0.824\pm0.034$ &$0.544\pm0.108$ \\ 
&$\beta$ &$-0.016\pm0.015$ &$0.054\pm0.019$ \\ 
&$R^2$ &$0.90$ &$0.31$  \\ 
\multirow{3}{*}{O} &$\alpha$ &$0.530\pm0.049$ &$0.464\pm0.123$ \\ 
&$\beta$ &$-0.019\pm0.016$ &$0.058\pm0.022$ \\ 
&$R^2$ &$0.79$ &$0.23$  \\ 
\multirow{3}{*}{Al} &$\alpha$ &$0.789\pm0.077$ &$1.096\pm0.088$ \\ 
&$\beta$ &$-0.061\pm0.028$ &$0.070\pm0.015$ \\ 
&$R^2$ &$0.82$ &$0.79$  \\ 
\multirow{3}{*}{Mg} &$\alpha$ &$0.556\pm0.062$ &$0.971\pm0.081$ \\ 
&$\beta$ &$0.042\pm0.023$ &$0.039\pm0.016$ \\ 
&$R^2$ &$0.62$ &$0.75$  \\ 
\multirow{3}{*}{Si} &$\alpha$ &$0.739\pm0.056$ &$0.878\pm0.059$ \\ 
&$\beta$ &$0.014\pm0.025$ &$0.017\pm0.011$ \\ 
&$R^2$ &$0.90$ &$0.85$  \\ 
\multirow{3}{*}{Ca} &$\alpha$ &$0.672\pm0.046$ &$1.022\pm0.072$ \\ 
&$\beta$ &$0.033\pm0.015$ &$0.031\pm0.014$ \\ 
&$R^2$ &$0.81$ &$0.83$  \\ 
\multirow{3}{*}{Na} &$\alpha$ &$0.987\pm0.202$ &$1.180\pm0.090$ \\ 
&$\beta$ &$-0.007\pm0.029$ &$0.018\pm0.016$ \\ 
&$R^2$ &$0.81$ &$0.81$  \\ 
\multirow{3}{*}{Ni} &$\alpha$ &$1.079\pm0.084$ &$1.200\pm0.071$ \\ 
&$\beta$ &$0.024\pm0.016$ &$-0.027\pm0.012$ \\ 
&$R^2$ &$0.94$ &$0.87$  \\ 
\multirow{3}{*}{Ti} &$\alpha$ &$1.510\pm0.277$ &$0.871\pm0.067$ \\ 
&$\beta$ &$0.081\pm0.072$ &$0.030\pm0.013$ \\ 
&$R^2$ &$0.71$ &$0.76$  \\ 

  \hline
	\end{tabular}

 \label{tab:coefs}

\end{table}

\begin{figure}[htbp!]
	\centering 
	\includegraphics[trim=0cm 0cm 0cm 0cm, scale=0.3, angle=0]{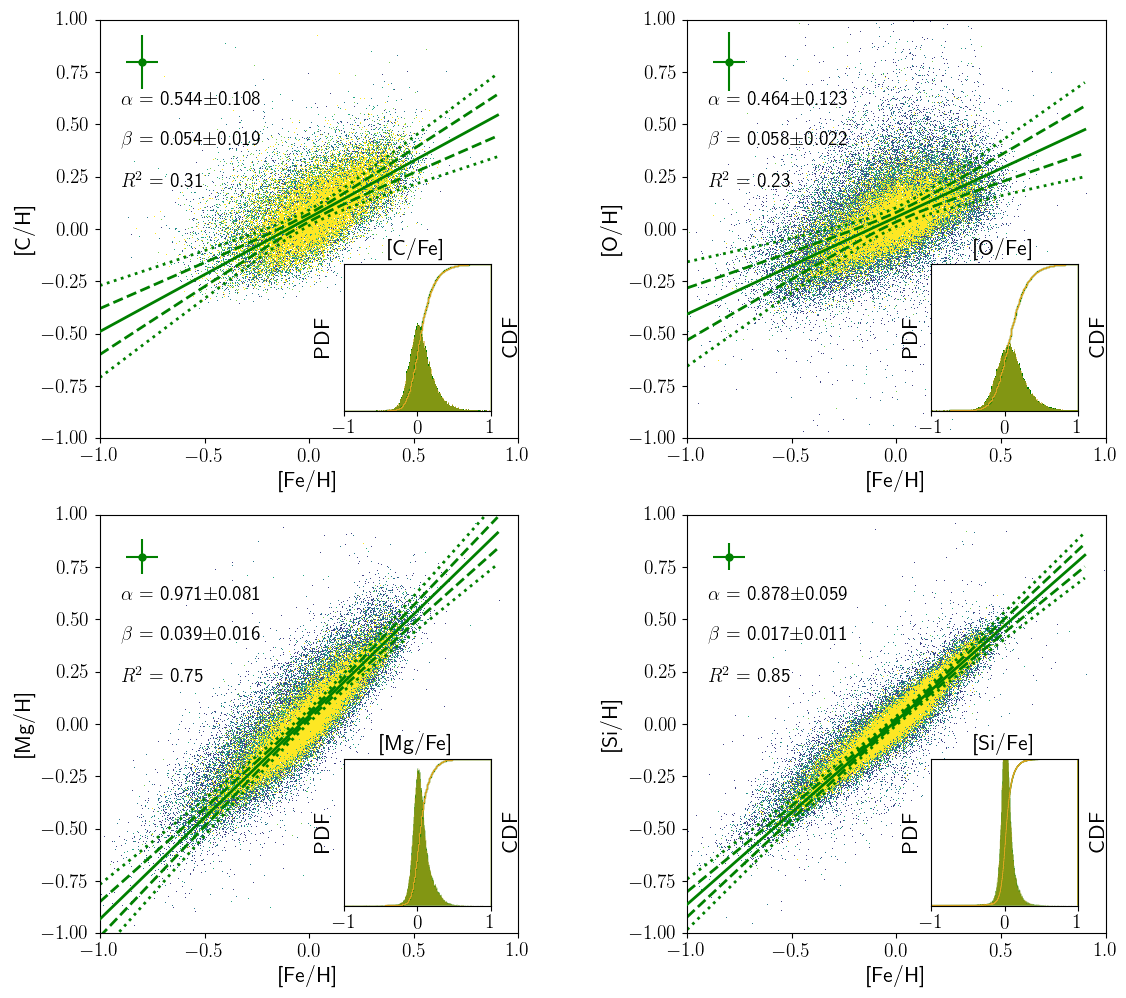}
	\caption{\edit{The [X/H]-[Fe/H] trends of (thin-disc) GK dwarfs, found by Orthogonal Distance Regressions (ODRs) following a resampling procedure (see text for  details). The model parameters (slope $\alpha$, intercept $\beta$, and coefficient of determination $R^2$) are shown in the individual panels. The underlying dot clouds  are the resampled GK dwarfs (nearly 100,000 data points after removing $\lesssim 4\%$ of outliers (Fig. \ref{fig:GK_ODRfits_coefs_outliers})). The typical error bars of the sample are shown in the upper left corner of each panel. The best-fits are shown in solid lines with their 1$\sigma$ and 2$\sigma$ error intervals shown in dash and dotted lines, respectively. The insets in individual panels show the  probability density functions (PDFs, y-axis scale omitted) and cumulative density functions (CDFs, y-axis scale omitted) of both the selected (original) sample of GK dwarfs (in green) and the resampled GK dwarfs (in yellow), which completely overlap with each other. Here we show only the cases for C, O, Mg, and Si; the full set of chemical trends for all elemental pairs can be found in Fig. \ref{fig:GK_resampled_ODRfits_ALL}.}}
	\label{fig:GK_resampled_ODRfits_COMgSi}
\end{figure}

\begin{figure}[htbp!]
	\centering 
	\includegraphics[trim=0cm 0cm 0cm 0cm, scale=0.3, angle=0]{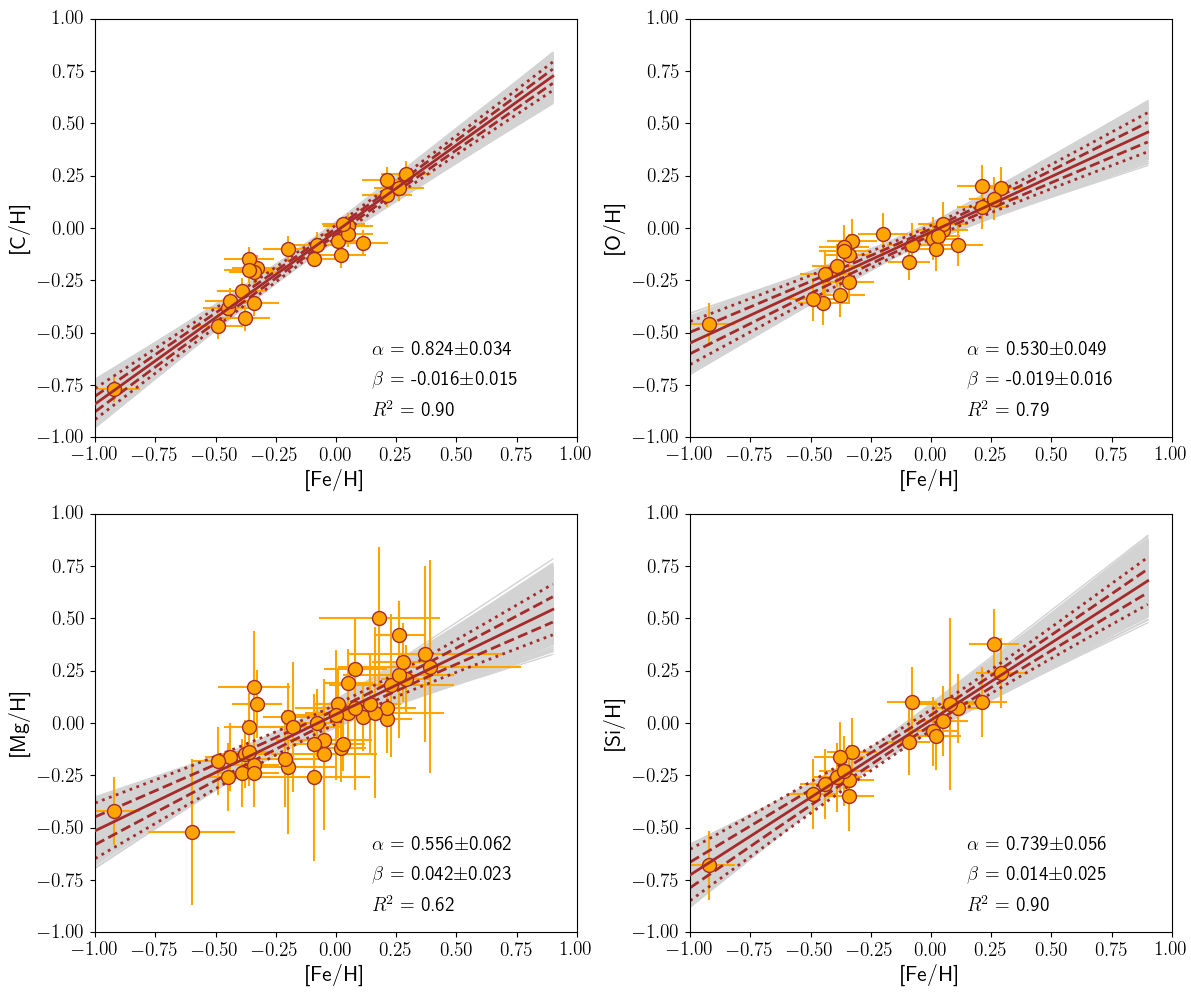}
	\caption{\edit{The [X/H]--[Fe/H] trends of M dwarfs found by ODRs following our bootstrap (with replacement) procedure (see text for details). The model parameters (slope $\alpha$, intercept $\beta$, and coefficient of determination $R^2$) are shown in the individual panels. In each panel the data points with error bars represent the selected (original) sample of M dwarfs; the lines in grey are the individual ODR fits for the individual sets of bootstrapped samples (with $\lesssim 4\%$ outliers removed; Fig. \ref{fig:Mboot_ODRfits_coefs_outliers}); the solid line in brown is the best-fit model, while the dashed and dotted lines indicate the 1$\sigma$ and 2$\sigma$ error intervals of the best-fit. Here we show only the cases for C, O, Mg, and Si; the full set of chemical trends for all elemental pairs can be found in Fig. \ref{fig:Mboot_ODRfits}.}}
	\label{fig:Mboot_ODRfits_COMgSi}
\end{figure}

\subsection{Chemical trends for GK dwarfs}
\label{sec:analysis_GK}

We adopt Orthogonal Distance Regression (ODR) for all linear regressions of chemical trends ([X/H]-[Fe/H]) in this work, and have considered the uncertainties in both [Fe/H] and [X/H]. The python package {\pyfont scipy.odr} \citep{2020SciPy-NMeth} has been employed to realise the regressions. 

\edit{Importantly, we have done a resampling process of our selected GK dwarfs before applying the regression analyses. Specifically, we have resampled the data in the [X/H]-[Fe/H] space for these GK dwarfs, with the size of each resampled subset fixed to 50 (i.e., comparable to the sample size of the subsequently analyzed M dwarfs). An ODR is applied to each resampled subset. We then repeat this resampling and regression process 2000 times, yielding 2000 regressions for each elemental pair. The purpose by going through this process is to make the regressions for the large sample of GK dwarfs comparable with those for the limited-size sample of M dwarfs. Due to the randomness of selection, this process will potentially yield outliers among all sets of regression results. We have identified and removed those outliers by employing a Mahalanobis-Distance outlier identification approach \citep{Ghorbani2019} at the two dimensional (2D) 3$\sigma$ level (0.9889; \citealt{Wang2015b}). The outliers account for less than $\sim 4\%$ of the 2000 sets of regressions in each case.}

\edit{The best-fit model parameters for each pair of [X/H]-[Fe/H] are the weighted averages of the final set of regression parameters (with the weights being the inverse squared 1$\sigma$ errors of individual regression parameters). The uncertainties of the best fits are the weighted standard deviations of these sets of regression parameters. The final results of the [X/H]-[Fe/H] trends for GK dwarfs are demonstrated in Fig. \ref{fig:GK_resampled_ODRfits_COMgSi} (for the cases of C, O, Mg, and Si), with the full set of comparisons for all elemental pairs in Fig. \ref{fig:GK_resampled_ODRfits_ALL}.}  

\subsection{Chemical trends for M dwarfs}
\label{sec:analysis_M}

Because of the limited size of the selected sample of M dwarfs (up to 43), we have employed the bootstrap technique (with replacement) to regenerate 2000 sets of samples. On each bootstrapped sample, an ODR is applied. This procedure also yields 2000 sets of regressions for M dwarfs for each pair of [X/H]-[Fe/H]. \edit{Similarly, due to the randomness of the bootstrap technique (with replacement), there are chances that only one or two data points are repeatedly sampled. This is particularly plausible in the case where the original sample is small. Therefore, we have also employed the Mahalanobis-Distance outlier identification approach \citep{Ghorbani2019} to identify and remove the outliers at the 2D 3$\sigma$ level \citep{Wang2015b}, as shown in Fig. \ref{fig:Mboot_ODRfits_coefs_outliers}. The number of those outliers is also found to be $\lesssim 4\%$ of the 2000 sets of regressions in each case.}

\edit{The same approach, adopted for finding the best-fit model parameters and their uncertainties for GK dwarfs, is also applied to find those best-fits and uncertainties for M dwarfs. The final results of the [X/H]-[Fe/H] trends for M dwarfs are shown in Fig. \ref{fig:Mboot_ODRfits_COMgSi} (for the cases of C, O, Mg, and Si), with the full set of comparisons for all elemental pairs in Fig. \ref{fig:Mboot_ODRfits}.}

\subsection{Consistency tests of chemical trends between GK and M dwarfs}

\begin{figure}[htbp!]
	\centering 
	\includegraphics[trim=0cm 0cm 0cm 0cm, scale=0.3, angle=0]{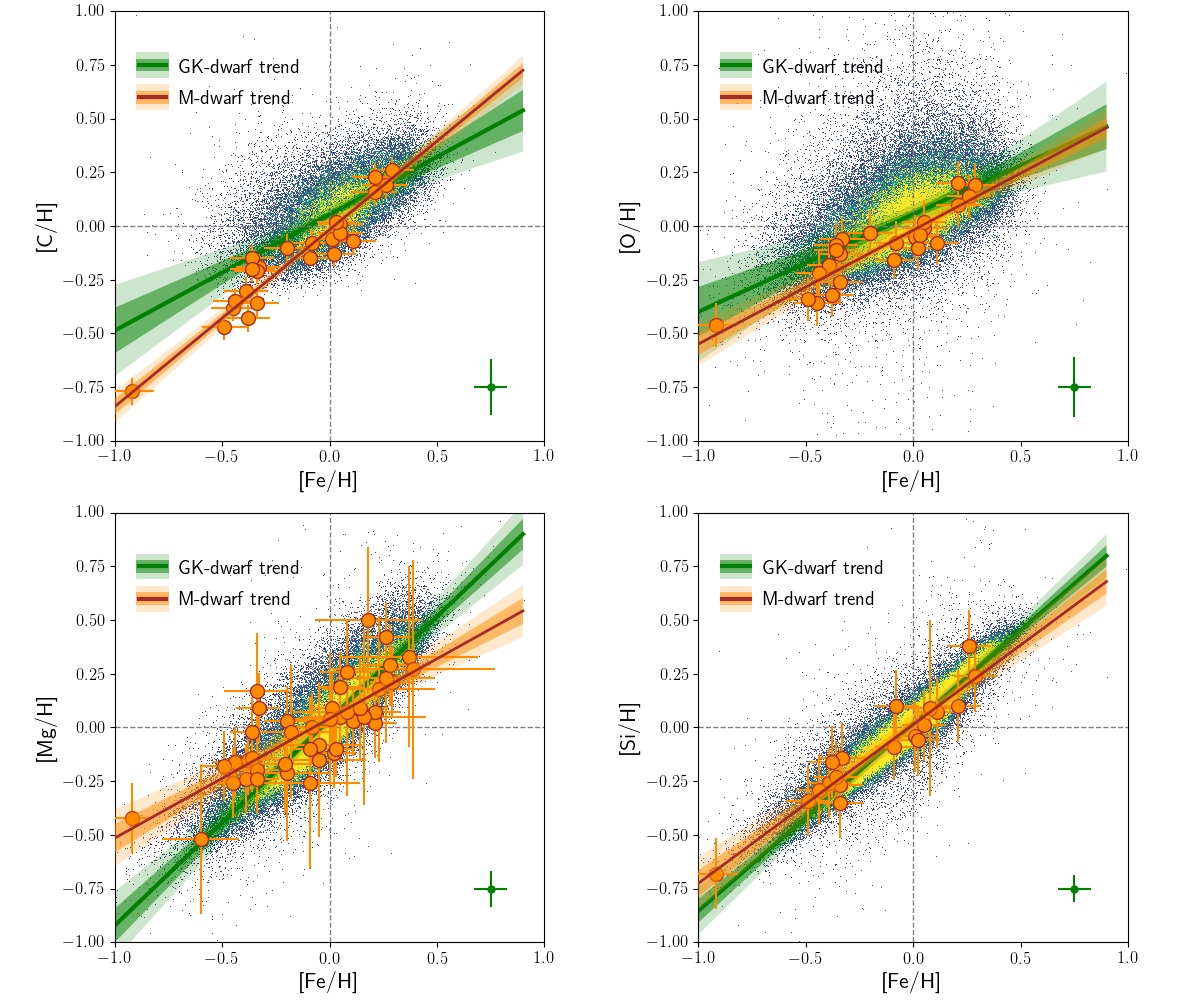}
	\caption{Comparison of the chemical trends of [X/H]-[Fe/H] between M dwarfs and GK dwarfs, \edit{based on Figs. \ref{fig:GK_resampled_ODRfits_COMgSi} and \ref{fig:Mboot_ODRfits_COMgSi}. The underlying colourful dots in density are the selected (original) sample of GK dwarfs, with their typical error bars shown in the lower right corner of each panel. The overlying orange dots (with error bars) are the selected (original) sample of M dwarfs.} The darker and lighter shaded wedges of individual regression lines represent their 1$\sigma$ and 2$\sigma$ \edit{error intervals}, respectively. \edit{Here we only show the cases for C, O, Mg, and Si, and the full set of comparisons of these chemical trends for all elemental pairs can be found in Fig. \ref{fig:DR3_M_scatter_withseldata_trends_ALL}.}}
	\label{fig:DR3_M_scatter_withseldata_trends_COMgSi}
\end{figure}

\begin{figure}[htbp!]
	\centering 
	\includegraphics[trim=0cm 0cm 0cm 0cm, scale=0.3, angle=0]{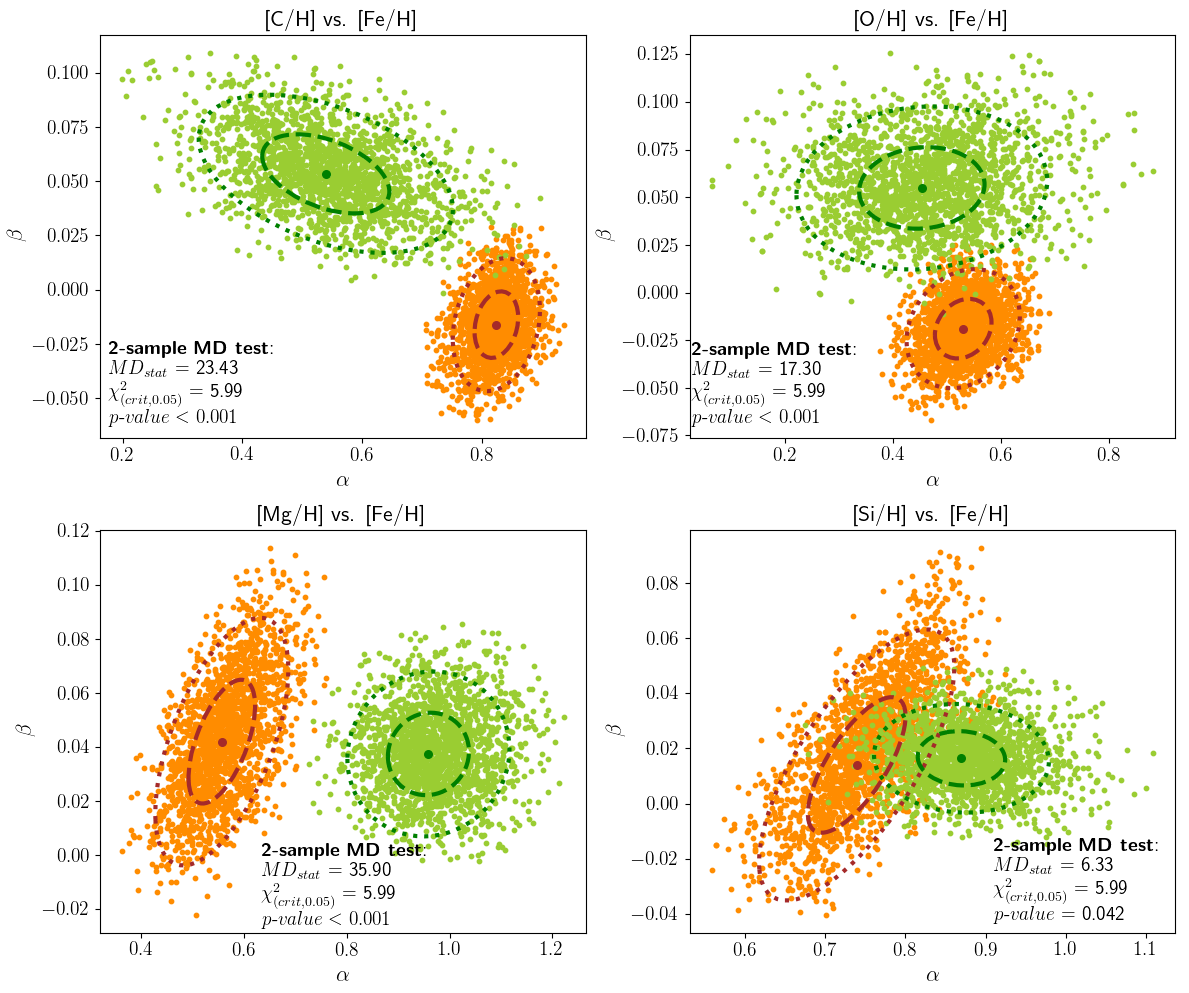}
	\caption{Comparison of the model \edit{coefficients} (slope $\alpha$ and intercept $\beta$) of the chemical trends between M dwarfs (orange dots) and GK dwarfs (green dots), \edit{along with the statistical test results. The dash and dotted ellipses indicate the 1$\sigma$ and 2$\sigma$ contours of their best-fit values (the dots in brown and in darker green). Here we show only the cases for C, O, Mg, and Si, and the full set of comparisons of these model coefficients for all elemental pairs can be found in Fig. \ref{fig:GK_Mboot_coefs_comp_ALL}.} See text for the details of the statistical test of the null hypothesis $H_0: \overline{(\alpha, \beta)_M} = \overline{(\alpha, \beta)_{GK}}$.}
	\label{fig:GK_Mboot_coefs_comp_COMgSi}
\end{figure}

\edit{The comparison of the chemical trends between M and GK dwarfs is shown in Fig. \ref{fig:DR3_M_scatter_withseldata_trends_COMgSi} for the cases of C, O, Mg, and Si and further in Fig. \ref{fig:DR3_M_scatter_withseldata_trends_ALL} for all elemental cases. The final regression results for both M and GK dwarfs are also summarised in Table \ref{tab:coefs}.}

\edit{Our statistical consistency tests are based on the final sets of model coefficients (slope $\alpha$ and intercept $\beta$; with uncertainties) of M and GK dwarfs, which we show in Fig. \ref{fig:GK_Mboot_coefs_comp_COMgSi} for the cases of C, O, Mg, and Si. The full set of comparisons for all elemental cases is shown in Fig. \ref{fig:GK_Mboot_coefs_comp_ALL}.}

Our null hypothesis is that there is no significant difference in the chemical trends of [X/H]-[Fe/H] between M and GK dwarfs. In other words, the joint model parameters ($\alpha$ and $\beta$) that determine those trends are statistically consistent in the 2D space:
$$
H_0: \overline{(\alpha, \beta)_M} = \overline{(\alpha, \beta)_{GK}}
$$

The null hypothesis can be first qualitatively examined based on the separation of the joint model parameters between the two samples, as shown in Fig. \ref{fig:GK_Mboot_coefs_comp_COMgSi}.  
\edit{For cases like C and Mg, the separation between the two data clouds of  model parameters is large enough to provide qualitative confidence that the null hypothesis may be rejected, whereas for cases like O and Si, a robust statistical test is necessary to come to a definitive conclusion.} 

\edit{For this, we adopt the Mahalanobis-Distance (MD) significance test \citep{Mahalanobis1930, Mardia1979} and apply it with the 2-sample, multivariate scenario (for the implementation see details in Appendix C). Notably, an MD test statistic ($MD_{stat}$) follows a $\chi^2$-distribution and can be compared to the critical value ($\chi^2_{crit, \alpha}$) at a chosen significance level (typically $\alpha=0.05$) of the $\chi^2$-distribution.} 

\edit{The test results, as shown in Fig. \ref{fig:GK_Mboot_coefs_comp_COMgSi}, allow us to conclude that for these four exemplary cases (i.e., C, O, Mg, and Si) the null hypotheses are all rejected, with the test statistic $MD_{stat}$ > $\chi^2_{crit, 0.05}$ and $p$-value \final{$<0.05$. However, the rejection for Si is marginal, with $p$-value = 0.042.}
As shown in Fig. \ref{fig:GK_Mboot_coefs_comp_ALL}, the null hypotheses for other elemental cases are also rejected, except for Na -- of which $MD_{stat} < \chi^2_{crit, 0.05}$ and $p$-value $> 0.05$.}

\section{Discussion}
\label{sec:discussion}

\subsection{The effect of different sample selections}\label{sec:discussion_sample}

\edit{Although we have been careful in selecting dwarf stars from the thin disc only to minimise the effects of varying stellar ages and birth environments, our analysis may still suffer from the different sampling regions for GK dwarfs (up to 2.4 kpc) and M dwarfs (up to 150 pc). To address this concern, we have done a supplementary analysis (with the same procedure) for the "nearby" sample of GK dwarfs and the (same) selected sample of M dwarfs (both confined to 150 pc from the Sun). The statistical test results (Fig. \ref{fig:GKnear_Mboot_coefs_comp_ALL}) suggest that our conclusions on rejection/non-rejection of the null hypothesis remain the same for all elements, except for Na -- for which $H_0$ is also rejected for the "nearby" scenario. This difference is mainly attributed to the larger deviation in $\beta$ of the fit for the "nearby" sample of GK dwarfs from that for the M dwarf sample. However, since the "nearby" GK dwarf sample is only $\lesssim$ 4\% of the full sample of GK dwarfs (as discussed in Section \ref{sec:data_GK}), we would be cautious to draw strong conclusions based on the outcome of the "nearby" scenario (alone).} 

\edit{Another concern with our sample selection may be that the abundance data of our selected samples of GK and M dwarfs come from different surveys. Indeed, although the same model atmosphere (MARCS) is adopted for abundance analysis for these surveys, discrepancies in data reduction, analysis pipelines, spectral line lists, and wavelength coverage will inevitably cause inherent differences in the resultant abundance data \citep{Hinkel2016}. As found in the \textit{Gaia}-ESO survey, differences in the stellar abundances analysed by different teams who adopt the exact same data and procedure are still unavoidable \citep{Smiljanic2014, Hourihane2023}. A cross-catalogue calibration of stellar abundances, as initiated for GALAH and APOGEE \citep{Nandakumar2022}, is an encouraging approach to minimise systematic differences between surveys, potentially resulting in a combined, more homogeneous dataset that would help refine such a statistical analysis. To partially address the in-homogeneity issue in the abundance data, we have carried out a supplementary analysis in Appendix \ref{app:apogee} with photospheric abundances of both GK and M dwarfs sourced from the same survey, APOGEE. The supplementary analysis results (Fig. \ref{fig:apo_GK_Mboot_coefs_comp_ALLL}) show that the inherent differences in the chemical trends between GK and M dwarfs remain, although the differences vary for different elements (e.g. Na and Ni). \final{Specifically, Na is rejected (with $p$-value = 0.021), while Ni is a clear case of non-rejection (with $p$-value = 0.992), contrary to the other two tests (Figs. B.9 and D.1).}}

\begin{figure}[htbp!]
	\centering 
	\includegraphics[trim=0cm 0cm 0cm 0cm, scale=0.6, angle=0]{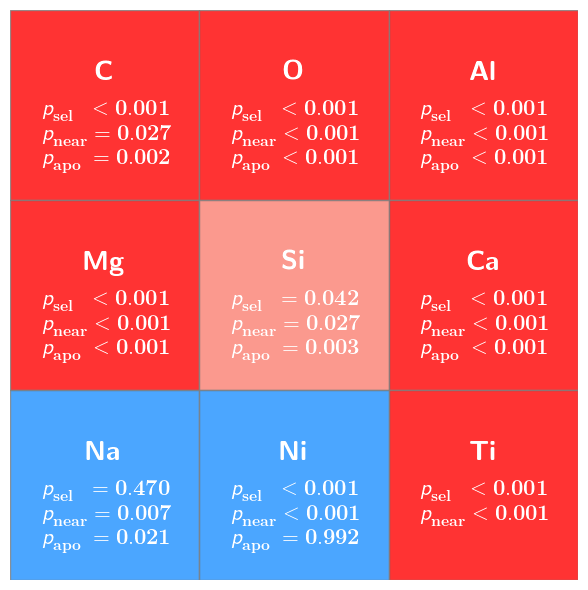} 
	\caption{Summary grid \edit{ of the \final{combined} test outcomes of the null hypothesis ($H_0$)} of the chemical trends [X/H]--[Fe/H] between M dwarfs and GK dwarfs for individual elements: red -- `rejection' of $H_0$ \final{(i.e., $p$-value $< 0.05$; while for the case of Si, it is distinguished in light red for its marginal rejection)}; blue -- `non-rejection/undetermined' of $H_0$ \final{(i.e., $p$-values are not consistently larger or smaller than 0.05 under different sample selections: `sel'-- the selected sample, `near'-- the "nearby" sample, and `apo'-- the APOGEE sample).}}
	\label{fig:GK_M_onsistency_panel}
\end{figure}

\edit{We summarise our combined test outcomes in a grid (Fig. \ref{fig:GK_M_onsistency_panel}), where different elemental cases shown on individual cells are coloured either in red -- representing the `rejection' of $H_0$ \final{(i.e., $p$-value $< 0.05$, as for the cases of C, O, Mg, Al, Ca, and Ti; while for the case of Si, it is distinguished in light red for its marginal rejection)} -- or in blue -- representing the `non-rejection/undetermined' of $H_0$ (in the cases of Na and Ni\final{, for which $p$-values are not consistently larger or smaller than 0.05 under different sample selections).}}

\subsection{Plausible reasons for the observed differences in [X/H]-[Fe/H] between M and GK dwarfs}
There are many \edit{plausible} reasons -- both physical and nonphysical -- for the observed differences in the chemical trends of [X/H]-[Fe/H] between M and GK dwarfs. The intrinsic differences between M and GK dwarfs can be best examined with a binary system that consists of a G/K dwarf as the primary and an M dwarf as its companion. This has been done in \cite{Ishikawa2020} for five such binary systems for eight individual elements including Na, Mg, Ca, Ti and Fe that are of interest to us. Their results show a typical discrepancy of $\leq$ 0.2 dex.
This discrepancy may be partially accounted for by atomic diffusion, which may act preferably over time to reduce the photospheric abundances of heavy elements relative to H ([X/H]) in the G/K primary relative to the M-dwarf companion \citep{Michaud2015, Souto2022}. Indeed, the effect of atomic diffusion in GK and M stars is mainly driven by gravitational settling, and the efficiency decreases with the depth of the convective envelope. The cooler the star, the deeper the convective envelope, and the less efficient the effect of atomic diffusion. In addition to this effect, the more time atomic diffusion has to act, the larger is the effect on the surface abundances. The magnitudes of the abundance changes caused by diffusion, varying by stellar mass and age, are found to be of the order of $\lesssim$ 0.1 dex for GK dwarfs \citep{Souto2022, Moedas2022}. By suppressing the diffusion effect approximately through examining the differences in [X/Fe] between GK primaries and M-dwarf companions with another set of binary systems, a mean abundance difference of $-0.05\pm0.03$ dex remains \citep{Souto2022}. 

Another factor that may contribute to the observed abundance differences between M and GK dwarfs is that the derived M-dwarf abundances are highly sensitive to $\log g$ because many spectral lines in the near-infrared regime are affected by pressure broadening under the high $\log g$ values of M dwarfs \citep{Ishikawa2020}. \final{Furthermore, while the effects of non-local thermaldynamic equilibrium (non-LTE) were considered in deriving the abundances of most elements of interest in the selected GK dwarf sample \citep{Buder2021}, such effects were not taken into account in the selected M dwarf sample \citep{Souto2017, Souto2018, Souto2022, Ishikawa2020, Ishikawa2022}. More studies on the effects of non-LTE in M dwarfs \citep[e.g.,][]{Olander2021} are required.}

\edit{Finally, it is unclear if there is any intrinsic difference in star formation and evolution that would cause the chemical deviations between Sun-like and cooler M-type dwarf stars. This is a question worth further investigating in the general context of stellar astrophysics including Galactic chemical evolution, asteroseismology, and star formation and evolution particularly for binary systems.}

\subsection{The potential impact on characterising planets}
\begin{figure}[htbp!]
	\centering 
	\includegraphics[trim=0cm 0cm 0cm 0cm, scale=0.6, angle=0]{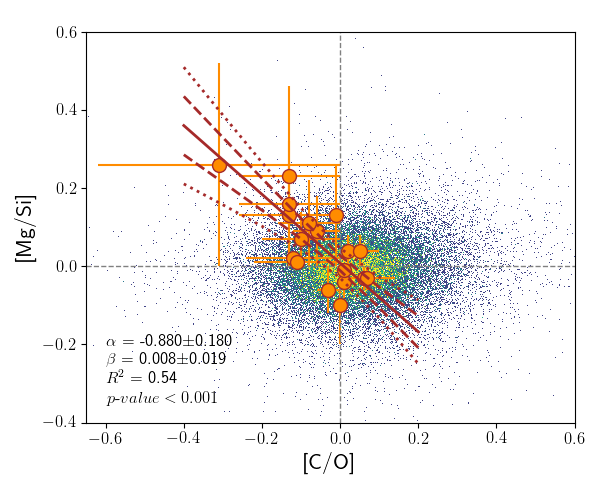} 
	\caption{The [Mg/Si]-[C/O] diagram for the M dwarf sample (overlying orange dots) and GK dwarf sample (underlying colourful dots in density). \edit{An ODR fit is applied to the M dwarf sample, with the best fit shown in the solid line and 1$\sigma$ and 2$\sigma$ error intervals shown in dash and dot lines (in brown), respectively. The regression parameters are also shown on the plot. A Student's t-test is also applied to test the significance of the slope different from zero, and the resultant $p$-value is less than 0.001.}}
	\label{fig:DR3_M_scatter_CO_MgSi}
\end{figure}

\edit{Since stellar chemical compositions have been shown to be crucial for characterising the chemical and internal properties of rocky exoplanets \citep{Hinkel2018, Putirka2019, Wang2019b, Wang2022, Wang2022b, spaargaren2023}, our observed differences for major rock-forming elements in the [X/H]-[Fe/H] space in M dwarfs relative to GK dwarfs will impact how the properties of M dwarf-hosted rocky planets may be interpreted.} For example, C/O and Mg/Si have been typically employed to indicate the propensity of planetary properties, e.g., silicate vs. carbide planets (lower vs. higher C/O) and pyroxene-rich vs. olivine-rich mantles (lower vs. higher Mg/Si) \citep[e.g.][]{Bond2010, Brewer2016, Wang2022, Wang2022b}. A qualitative comparison of the M and GK dwarfs on a [Mg/Si]--[C/O] diagram (Fig. \ref{fig:DR3_M_scatter_CO_MgSi}) suggests that [Mg/Si] of M dwarfs appear to decrease with the increase of [C/O] while there is no obvious trend between [Mg/Si] and [C/O] of GK dwarfs. \edit{Indeed, a simple Student's t-test shows that the slope found for M dwarfs in the [Mg/Si]--[C/O] space is significantly different from zero (with $p$-$value <0.001$). Concerning the plausible sensitivity of those two data points with [Mg/Si] > \final{0.2 dex} to the test outcome, we have also done a \final{sensitivity} test by removing those two data points and the result shows that the slope of the new trend for M dwarfs remains statistically significant (Fig. \ref{fig:DR3_M_scatter_CO_MgSi_sens_test}).} This discrepancy \edit{in the [Mg/Si]--[C/O] space between M and GK dwarfs} is consequent from the observed differences in the individual [X/H]--[Fe/H] trends shown in Fig. \ref{fig:DR3_M_scatter_withseldata_trends_COMgSi}. A further analysis of this discrepancy and its impact on constraining planetary properties will be carried out in our subsequent work on the nebular condensation of \edit{various types of stars} (Wang et al., in prep).

\section{Conclusions}\label{sec:conclusion}
\edit{We have carried out a thorough analysis of the chemical compositions of well-selected samples of M and GK dwarfs, which yields statistically constrained [X/H]-[Fe/H] trends for both samples. We further test the null hypothesis that these trends between M and GK dwarfs are indistinguishable. We find that this null hypothesis is rejected for all elements, except for Na (based on our selected samples). In considering different sample selections, the rejection of the null hypothesis \final{strongly} holds for C, O, Mg, Al, Ca, and Ti\final{, while the rejection for Si is marginal. These elements}, along with Fe, are crucial \final{for constraining} the detailed properties of rocky planets or planetary building blocks \citep{Wang2019b, Bitsch2020, Timmermann2023}. We show the potential impact of this difference to first order by analysing the [Mg/Si]-[C/O] ratios. We note that the test outcomes for Na and Ni are not always consistent between different sample selections, warranting future investigations.}  

Nonetheless, based on these observed differences, a blind application of the chemical abundances of Sun-like stars to constrain those of cooler, M dwarf stars (other than those as companions in binary systems) has to be cautioned. Regardless of the root-causes of these observed differences, an investigation of dedicated modelling techniques for M dwarf atmospheres, as well as an increasing set of benchmark, homogeneous abundance analyses, are recommended. 

In the meantime, the statistically constrained [X/H]-[Fe/H] trends based on the best available M dwarf abundances (summarised in Table \ref{tab:coefs}) are of practical use to constrain the chemical abundances of M dwarfs given available [Fe/H] measurements. This will help towards an improved understanding of the composition and other properties of rocky worlds that are found in increasing number around M dwarfs.

\begin{acknowledgements}
\edit{We thank the referee Ulrike Heiter for a constructive review, which has significantly improved the quality of the manuscript.} 
We acknowledge useful discussion with Sven Buder, Fan Liu and David Yong about the GALAH sample as well as Diogo Souto about the APOGEE sample. 
The statistical analysis in this work also benefits from discussion with statisticians including G. Jogesh Babu at Penn State and Lukas Graz, Emily Garvin and Markus Bonse at ETH Zurich. We also thank Paolo Sossi, Caroline Dorn and Johanna Teske for useful discussion. This work has been carried out within the framework of the National Centre of Competence in Research PlanetS supported by the Swiss National Science Foundation (SNSF) under grants 51NF40\_182901 and 51NF40\_205606. H.S.W and S.P.Q acknowledge the financial support of the SNSF. 
\end{acknowledgements}

%
%
\bibliographystyle{aa}
\bibliography{ref}

\clearpage
\onecolumn
\appendix

\section{Population classification of the sample of GK dwarfs}\label{app:class}

\begin{figure*}[htbp!]
	\centering 
	\begin{subfigure}[b]{0.48\textwidth}
	\includegraphics[width=\textwidth]{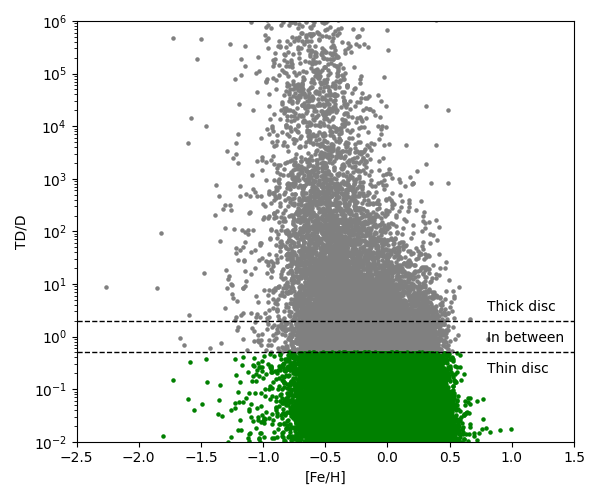} 
	\end{subfigure}
   ~
	\begin{subfigure}[b]{0.48\textwidth}
	\includegraphics[width=\textwidth]{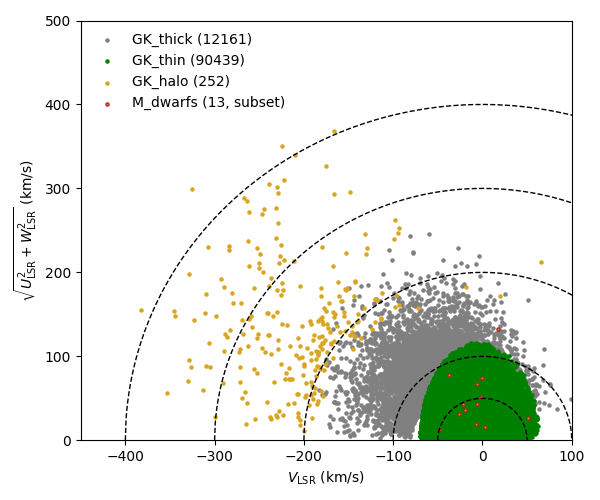} 
    \end{subfigure}
	~    
	\caption{\edit{Population classification of the sample of GK dwarfs.} Left: Classification of thin- and thick-disc stars based on a probability ratio (TD/D) defined in \cite{Bensby2014} (here, the "in-between" stars are grouped into "thick-disc" stars). Right: The Toomre diagram of the sample of GK dwarfs based on their kinematic data (\edit{$U_{\mathrm{LSR}}$, $V_{\mathrm{LSR}}$, and $W_{\mathrm{LSR}}$, relative to the local standard of rest (LSR)}; \citealt{Bensby2014}). \edit{The dash curves indicate the isotropic total space velocities ($v_{tot}$).} A subset sample of M dwarfs with available kinematic data \citep{Ishikawa2022} are also shown for comparison.}
	\label{fig:GK_M_sel}
\end{figure*}

\edit{We follow the kinematic selection criteria of \cite{Bensby2014} (for details see their Appendix A) to classify our sample of GK dwarfs (102,852 in total) into potential thin-disc stars, thick-disc stars, and halo stars. The kinematic data in the heliocentric velocity frames ($U$, $V$, $W$) of our sample are obtained from the Value-Added-Catalogue of GALAH DR3 \citep{Buder2021}, and these data are then converted to the local standard of rest (LSR) velocity frames ($U_{\mathrm{LSR}}$, $V_{\mathrm{LSR}}$, $W_{\mathrm{LSR}}$) by adding the peculiar Solar velocities (relative to the LSR), $(U_\odot, V_\odot, W_\odot) = (11.10, 12.24, 7.25)$ km/s from \cite{Schonrich2010} as adopted in the Bensby criteria. It is noteworthy that the adopted $V_\odot$ value is relatively lower than the GALAH intrinsic $V_\odot$ value (15.17 km/s), but to keep consistency with the Bensby criteria, we opt for the same set of peculiar Solar velocities (i.e., \citealt{Schonrich2010}). Then, we calculate the probability ratios of thick disc-to-thin disc (TD/D) and thick disc-to-halo (TD/H) memberships, as defined in the Appendix A of \cite{Bensby2014}. According to the Bensby criteria, we have 90,439 GK dwarfs with thin-disc kinematics ($TD/D < 0.5$), 7,034 GK dwarfs with thick-disc kinematics ($TD/D < 2$), and 252 GK dwarfs with halo kinematics ($TD/H < 1$). There are also 5,127 GK dwarfs with kinematics that are in between thin-disc and thick-disc values ($0.5<=TD/D <= 2$). For simplicity, we group these "in-between" GK dwarfs into thick-disc stars, the total number of which are therefore 12,161. We demonstrate the populations of these GK dwarfs in Fig. \ref{fig:GK_M_sel} (left panel: $TD/D$ vs. [Fe/H]; right panel: a Toomre diagram). For comparison, a subset sample of M dwarfs with available kinematic data \citep{Ishikawa2022} are also shown on the Toomre diagram. To clarify, a Toomre diagram can also be used to classify stellar populations, but here we use it only for visualisation and our classification is purely based on the kinematic selection criteria of \cite{Bensby2014}. To first approximation, the total space velocities $v_{tot} \equiv \sqrt{U_{\mathrm{LSR}}^2+V_{\mathrm{LSR}}^2+W_{\mathrm{LSR}}^2}$ for our thin-disc, thick-disc, and halo GK dwarfs are $\lesssim 100$ km/s, $\sim$ 100--200 km/s, and $\gtrsim 200$ km/s (the boundaries are qualitative).
}

\section{Full sets of comparisons and analyses in various cases}\label{app:full_compar}
\begin{figure*}[htbp!]
	\centering 
	\includegraphics[trim=0cm 0cm 0cm 0cm, scale=0.4, angle=0]{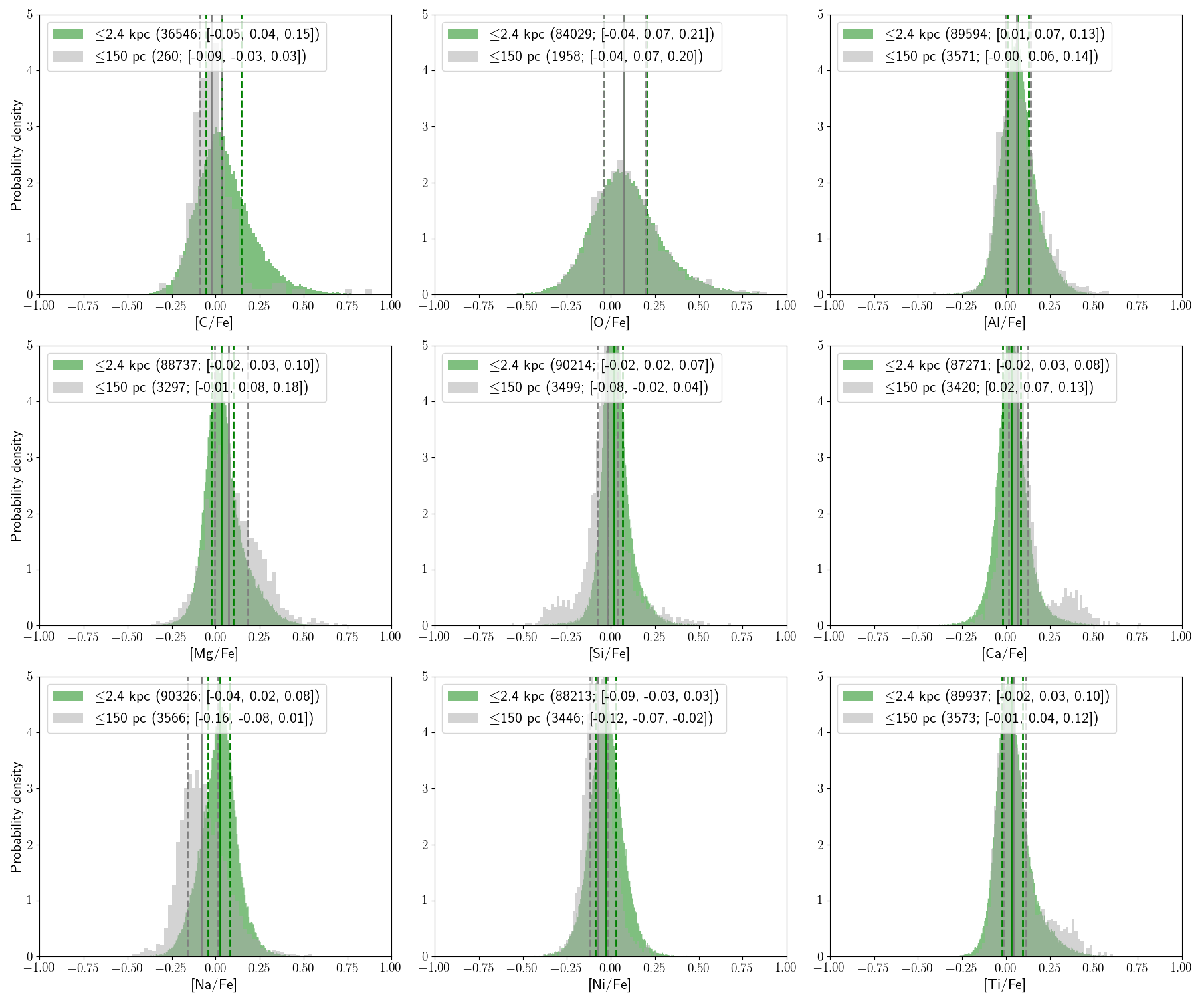} 
	\caption{\edit{Same as Fig. \ref{fig:DR3_data_hist_COMgSi}, but for the full set of comparisons of the probability densities of [X/Fe] between the full ("$\leq$ 2.4 kpc") and nearby ("$\leq$ 150 pc") samples of thin-disc GK dwarfs for these nine pairs of rock-forming elements.}}
	\label{fig:DR3_data_hist_main}
\end{figure*}

\begin{figure*}[htbp!]
	\centering 
	\includegraphics[trim=0cm 0cm 0cm 0cm, scale=0.4, angle=0]{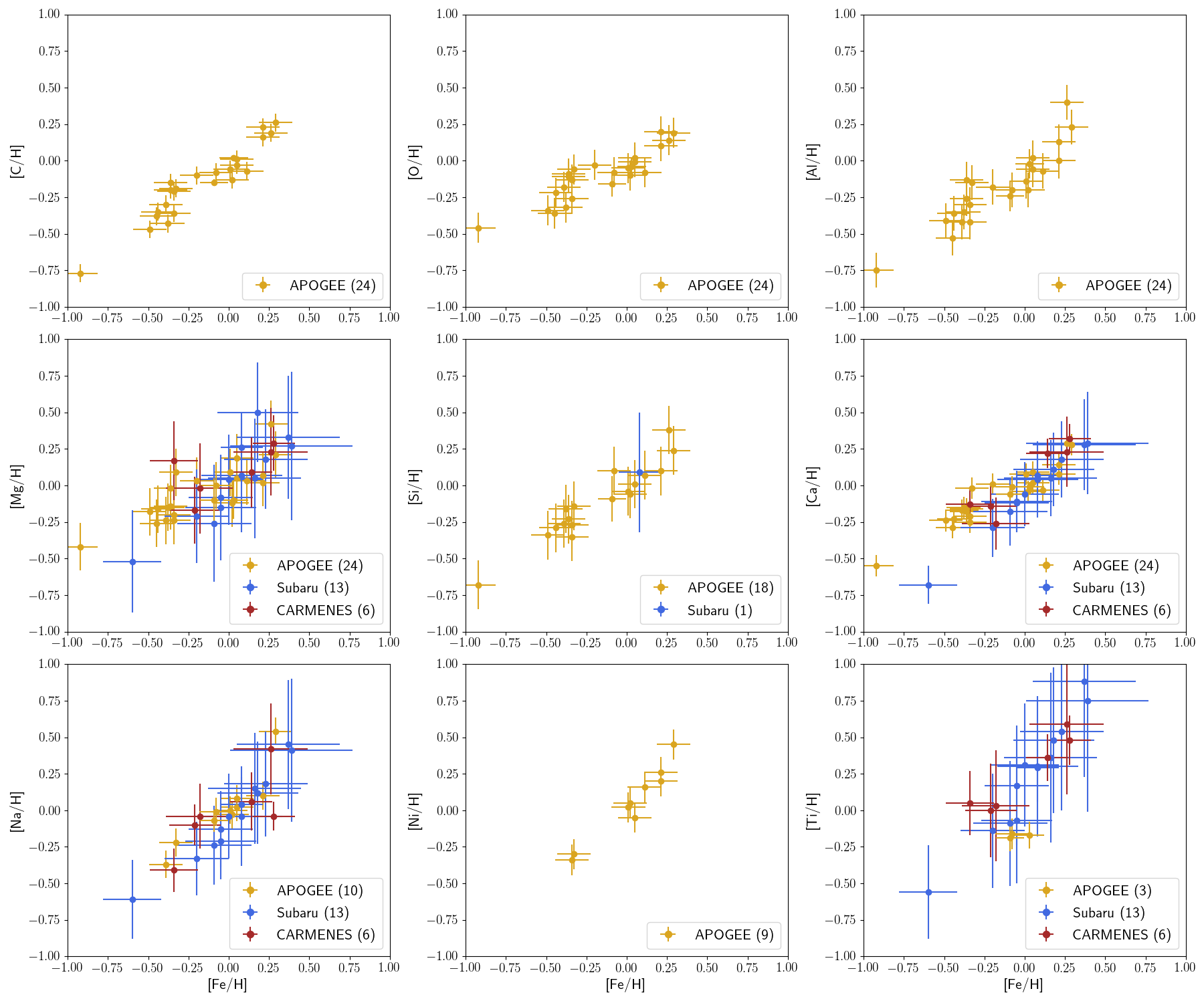}
	\caption{\edit{Same as Fig. \ref{fig:M_data_scatter_werr_COMgSi}, but for the full set of selected M-dwarf abundance data for these nine pairs of rock-forming elements.}}
	\label{fig:M_data_scatter_werr}
\end{figure*}

\begin{figure*}[htbp!]
	\centering 
	\includegraphics[trim=0cm 0cm 0cm 0cm, scale=0.4, angle=0]{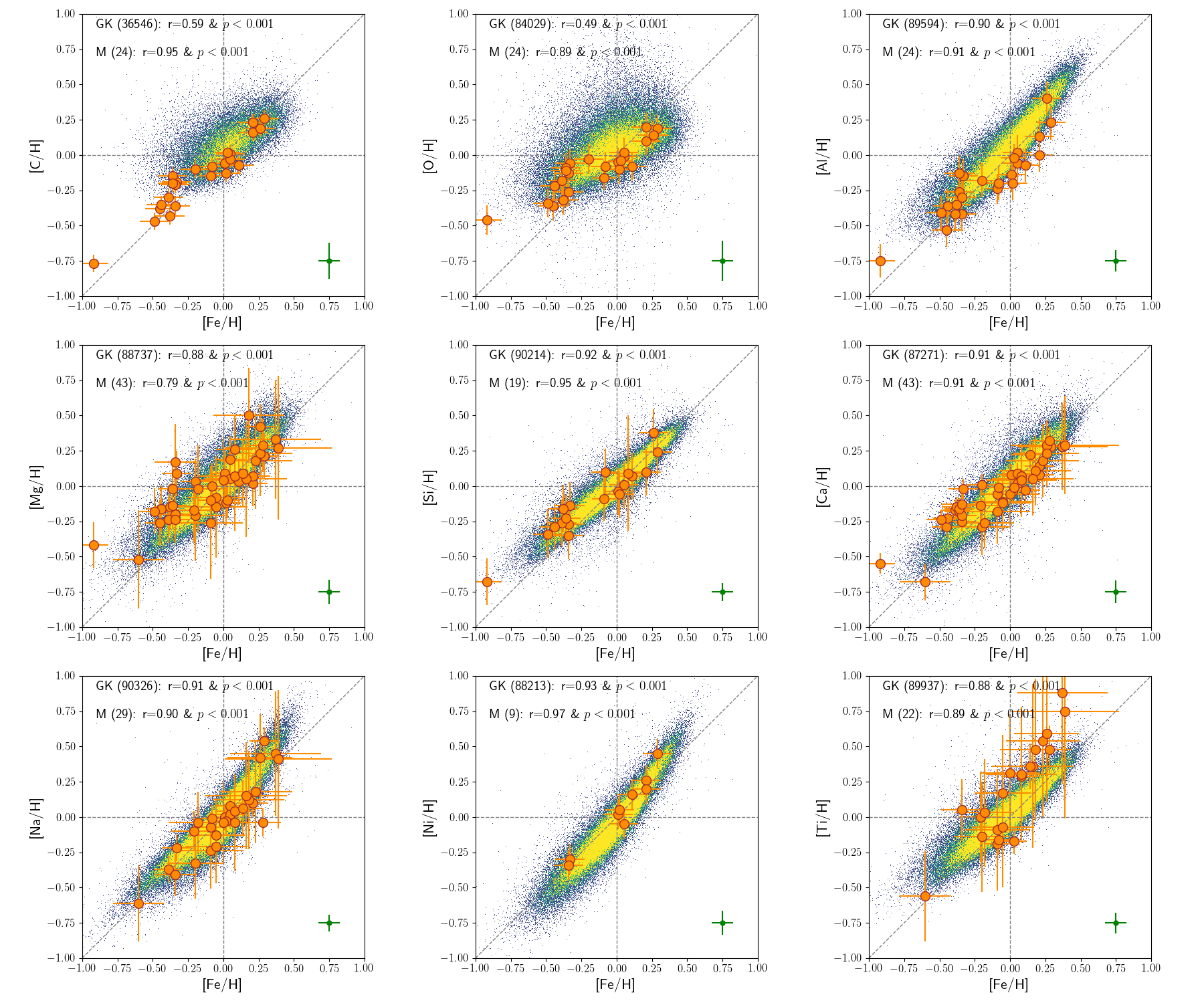}
	\caption{\edit{Same as Fig. \ref{fig:DR3_M_scatter_withseldata_COMgSi}, but for the full set of comparisons of the selected samples of GK and M dwarfs in the [X/H]-[Fe/H] space for these nine pairs of rock-forming elements.}}
	\label{fig:DR3_M_scatter_withseldata}
\end{figure*}

\begin{figure*}[htbp!]
	\centering 
	\includegraphics[trim=0cm 0cm 0cm 0cm, scale=0.4, angle=0]{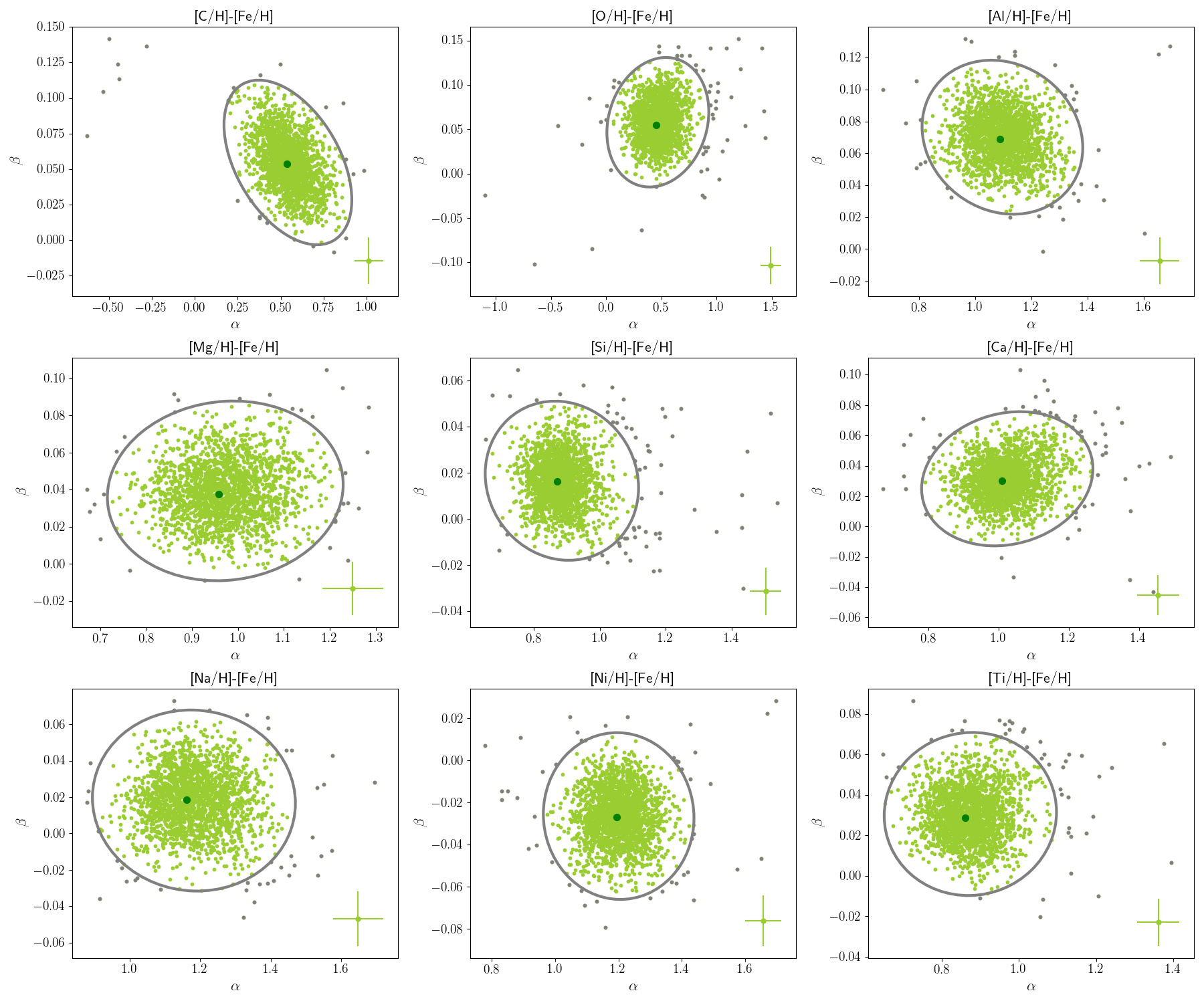}
	\caption{\edit{The distribution of the (raw) 2000 sets of model parameters (slope $\alpha$ and intercept $\beta$) of GK dwarfs for each pair of [X/H]--[Fe/H], following a resampling and regression process (see the details in Sect. \ref{sec:analysis_GK}). The typical error bars of these individual model parameters (in green) are indicated in the lower right corner of each panel, with the best-fit values shown in darker green. The few dots in grey are the outliers identified at the 2D 3$\sigma$ level (ellipses in grey) with a Mahalanobis-Distance outlier identification approach \citep{Ghorbani2019}. Those outliers account for $\lesssim 4\%$ of the raw sets of model parameters and haven been removed for further analysis.}}
	\label{fig:GK_ODRfits_coefs_outliers}
\end{figure*}

\begin{figure*}[htbp!]
	\centering 
	\includegraphics[trim=0cm 0cm 0cm 0cm, scale=0.4, angle=0]{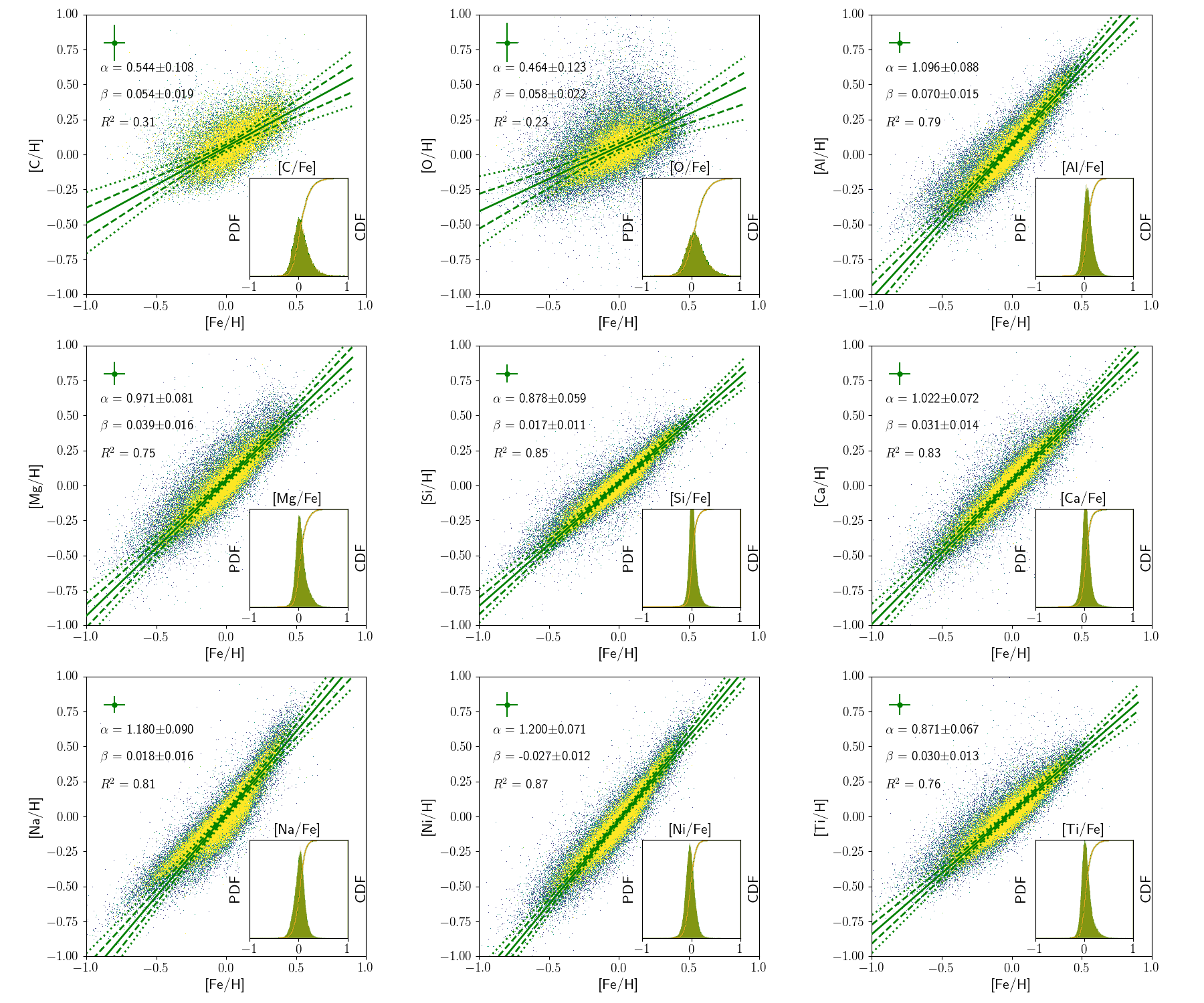}
    \caption{\edit{Same as Fig. \ref{fig:GK_resampled_ODRfits_COMgSi}, but for the full set of chemical trends of GK dwarfs for these nine pairs of rock-forming elements.}}	
	\label{fig:GK_resampled_ODRfits_ALL}
\end{figure*}

\begin{figure*}[htbp!]
	\centering 
	\includegraphics[trim=0cm 0cm 0cm 0cm, scale=0.4, angle=0]{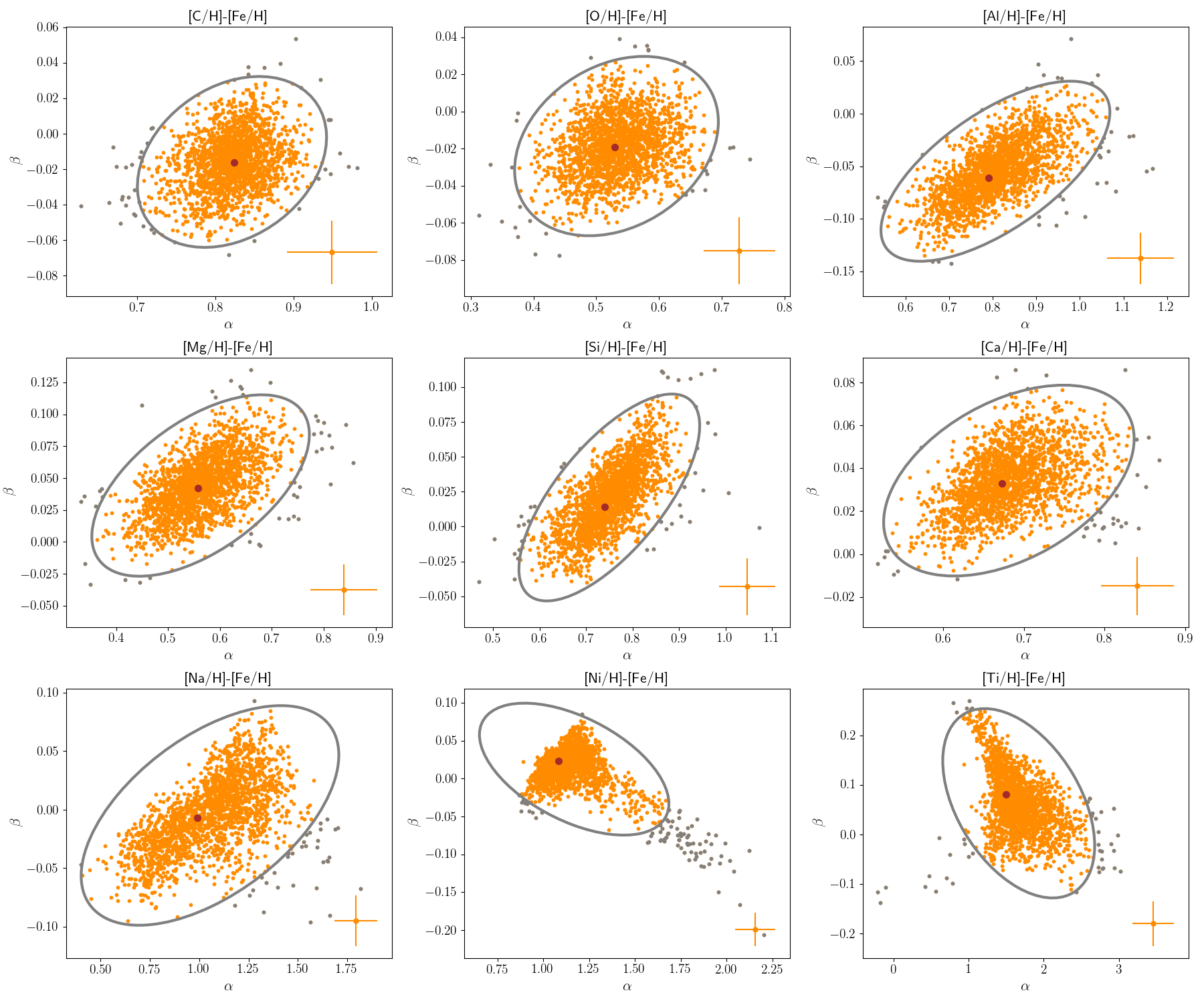}
	\caption{\edit{The distribution of the (raw) 2000 sets of model parameters (slope $\alpha$ and intercept $\beta$) of M dwarfs for each pair of [X/H]--[Fe/H], following a bootstrap and regression process (see the details in Sect. \ref{sec:analysis_M}). The typical error bars of these individual model parameters (in orange) are indicated in the lower right corner of each panel, with the best-fit values shown in brown. The few dots in grey are the outliers identified at the 2D 3$\sigma$ level (ellipses in grey) with a Mahalanobis-Distance outlier identification approach \citep{Ghorbani2019}. Those outliers account for $\lesssim 4\%$ of the raw sets of model parameters and haven been removed for further analysis.}}
	\label{fig:Mboot_ODRfits_coefs_outliers}
\end{figure*}

\begin{figure*}[htbp!]
	\centering 
	\includegraphics[trim=0cm 0cm 0cm 0cm, scale=0.4, angle=0]{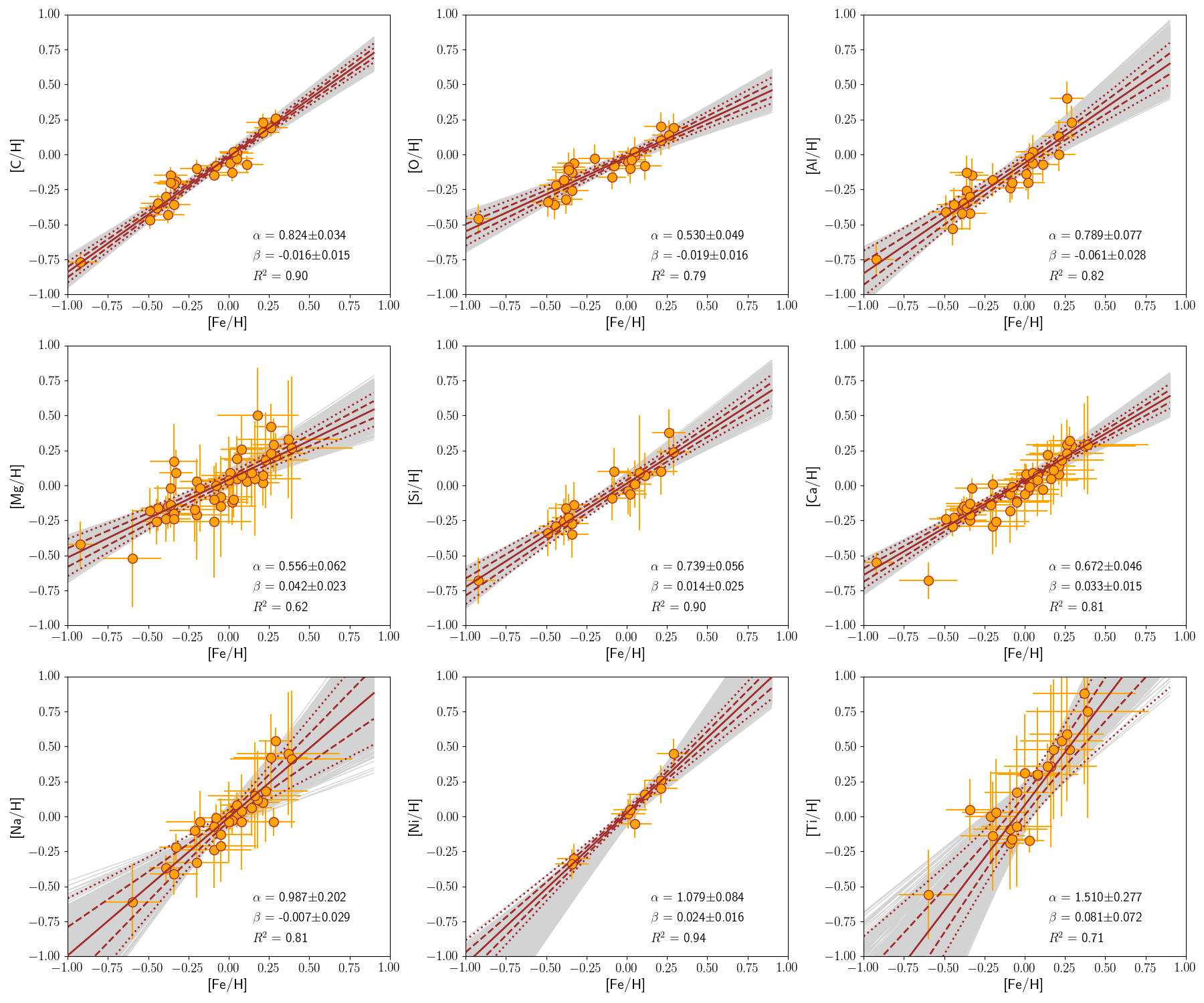}
	\caption{\edit{Same as Fig. \ref{fig:Mboot_ODRfits_COMgSi}, but for the full set of chemical trends of M dwarfs for these nine pairs of rock-forming elements.}}	
	\label{fig:Mboot_ODRfits}
\end{figure*}

\begin{figure*}[htbp!]
	\centering 
	\includegraphics[trim=0cm 0cm 0cm 0cm, scale=0.4, angle=0]{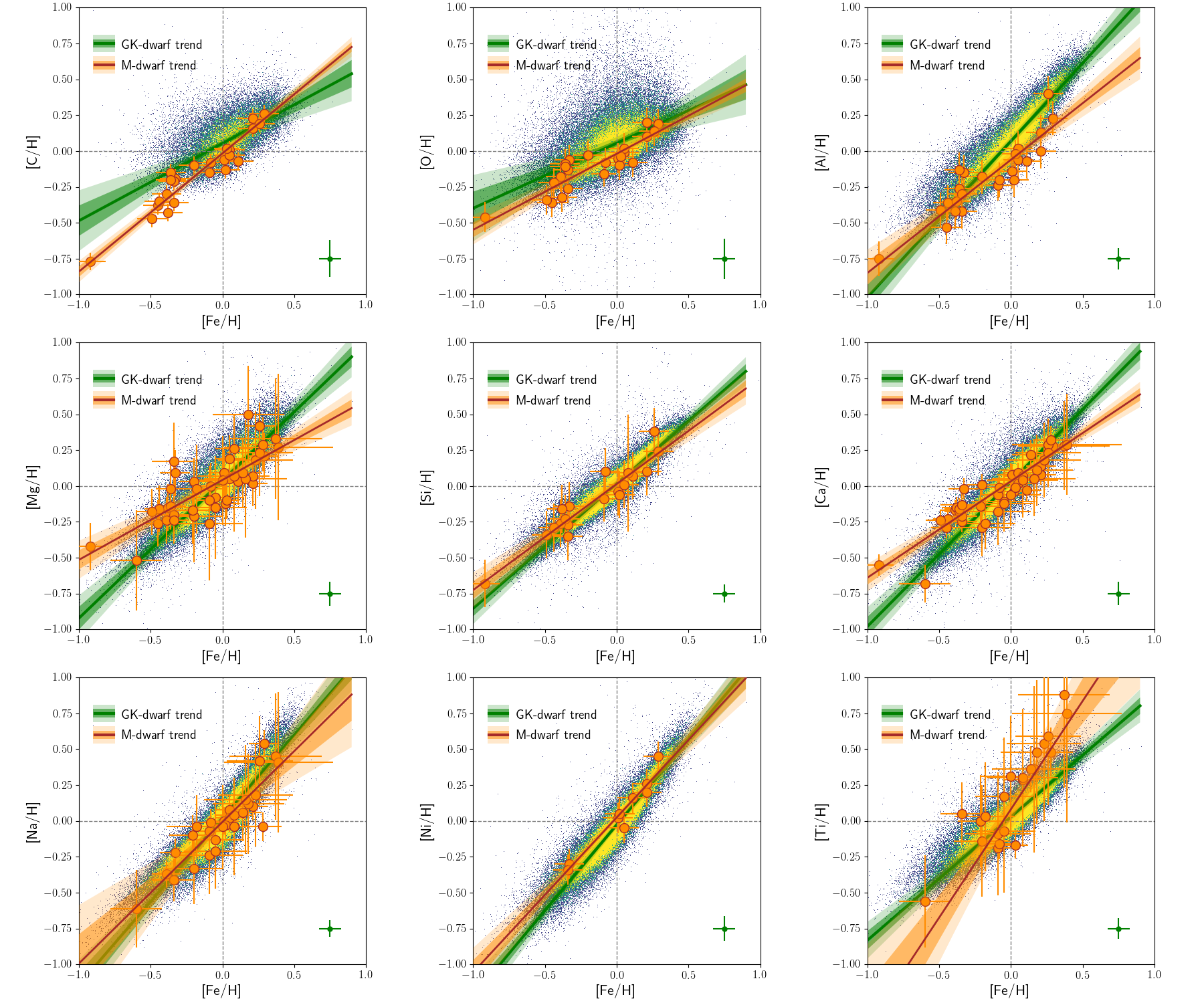}
	\caption{Same as Fig. \ref{fig:DR3_M_scatter_withseldata_trends_COMgSi}, but for the full set of comparisons of the chemical trends between M and GK dwarfs for these nine pairs of rock-forming elements.} 
	\label{fig:DR3_M_scatter_withseldata_trends_ALL}
\end{figure*}

\begin{figure*}[htbp!]
	\centering 
	\includegraphics[trim=0cm 0cm 0cm 0cm, scale=0.4, angle=0]{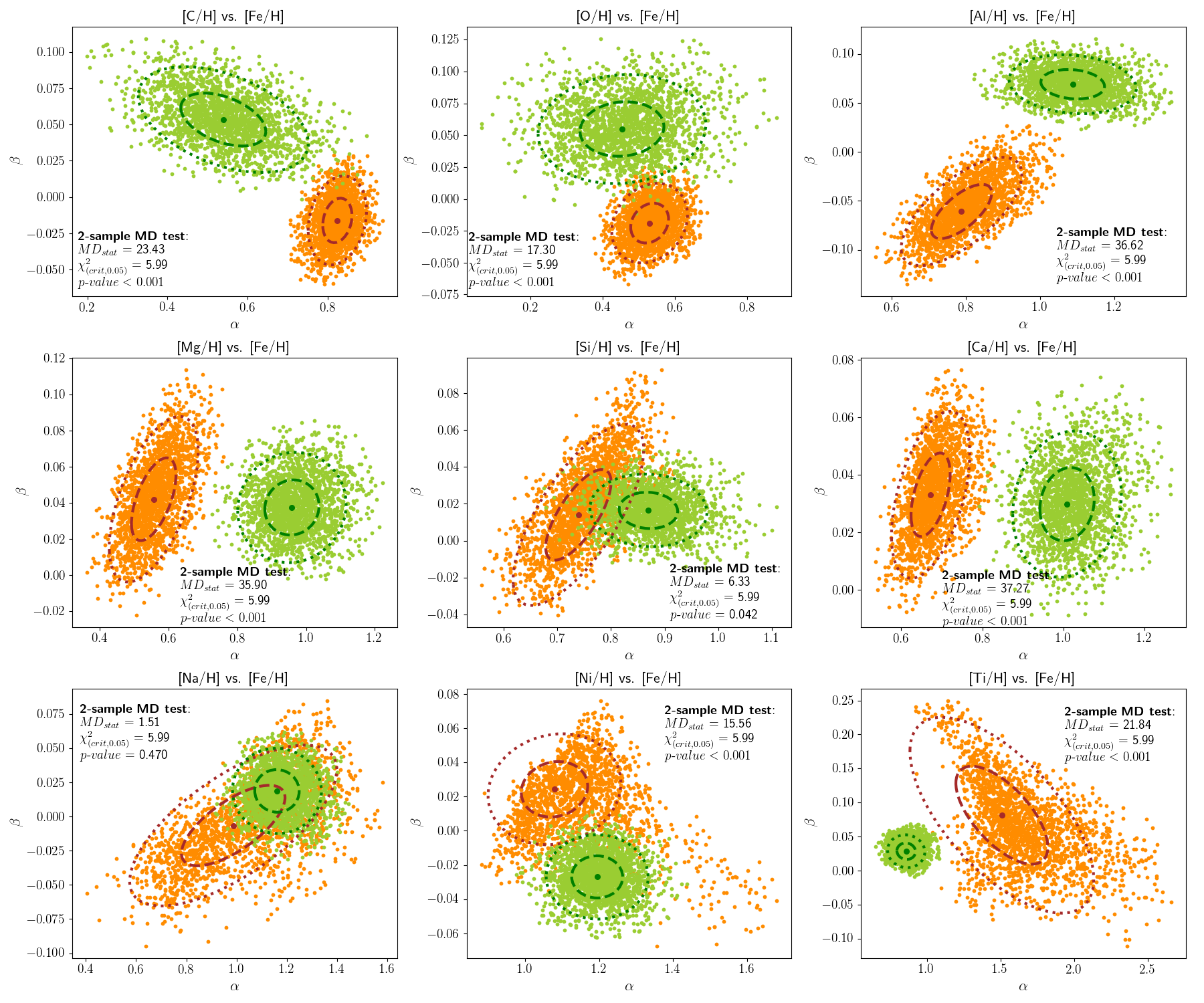}
	\caption{Same as Fig. \ref{fig:GK_Mboot_coefs_comp_COMgSi}, but for the full set of comparisons of the model coefficients between M dwarfs (orange dots) and GK dwarfs (green dots) for these nine pairs of rock-forming elements.} 
	\label{fig:GK_Mboot_coefs_comp_ALL}
\end{figure*}

\edit{This section shows the full sets of comparisons and analyses done for all of these nine pairs of rock-forming elements (Fig.s \ref{fig:DR3_data_hist_main}--\ref{fig:GK_Mboot_coefs_comp_ALL}). Except Figs. \ref{fig:GK_ODRfits_coefs_outliers} and \ref{fig:Mboot_ODRfits_coefs_outliers}, the subsets (for cases of C, O, Mg, and Si) of these nine-panel plots have been presented in the main body of the manuscript.}

\clearpage

\section{The implementation of a 2-sample multivariate Mahalanobis Distance significance test}\label{app:mahala}
\edit{A multivariate Mahalanobis Distance significance test between two samples with (multivariate) means \vec{\overline{x}_1} and \vec{\overline{x}_2} and common variance-covariance matrix \vec{S} is defined by the following formula \citep{Mahalanobis1930, Mardia1979}: 
\begin{equation}
D^2 = (\overline{x}_1 - \overline{x}_2)' S^{-1} (\overline{x}_1 - \overline{x}_2)
\end{equation}
where $D^2$ is the squared Mahalanobis Distance (i.e. the test statistic, $MD_{stat}$), and $S = ((n_1-1)S_1 + (n_2-1)S_2) / (n_1 + n_2 -2)$ is pooled from the individual variance-covariance matrices ($S_1$ and $S_2$) of the two samples of sizes, $n_1$ and $n_2$. This is the unbiased pooling of $S_1$ and $S_2$, weighted by their individual sample sizes minus 1.} 

\edit{Please note that the sample means can be either weighted or unweighted, so be the variance-covariance matrices. Here, we have adopted the weighted averages as the sample means, with the weights ($w$) being the inverse squared errors ($\sigma$) in the multivariate data of the individual samples, i.e., $w_{ki} = 1/\sigma_{ki}^2$, where $k$ is the $k$-th variable of the $p$-dimensional/variate data ($p=2$ as accounted for slope and intercept in this case) and i is $i$-th data point of an individual sample. In calculating the weighted variance-covariance matrix of an individual sample, the jointed, quadratic weights ($w_q$) of the multivariate data are adopted, i.e, $w_{qi} = \sqrt{\sum_{k=1}^p w_{ki}^2}$.} 

\edit{The test statistic ($MD_{stat}$) follows a $\chi^2$ distribution with $p$ degrees of freedom. The null hypothesis test is carried out by comparing $MD_{stat}$ with the critical value ($\chi^2_{crit, \alpha}$) of the $\chi^2$ distribution at a given significance level ($\alpha$, typically chosen at 0.05). If $MD_{stat} \geq \chi^2_{crit,\alpha}$ and the corresponding $p$-$value$ (i.e., the probability of observing the tested or more extreme data under the assumption that the null hypothesis is true) equal to or smaller than $\alpha$, the null hypothesis is rejected. Otherwise, namely, $MD_{stat} < \chi^2_{crit,\alpha}$ and $p$-$value$ > 0.05, the null hypothesis is not rejected as of insufficient evidence to do so.} 

\edit{We have made our Python script\footnote{\url{https://github.com/astro-seanwhy/AstroStats}} in implementing the test method as described open-accessed for the reader. } 

\clearpage
\section{A supplementary analysis for the "nearby" samples of both GK and M dwarfs (within 150 pc)}\label{app:150pc}

\begin{figure*}[htbp!]
	\centering 
	\includegraphics[trim=0cm 0cm 0cm 0cm, scale=0.4, angle=0]{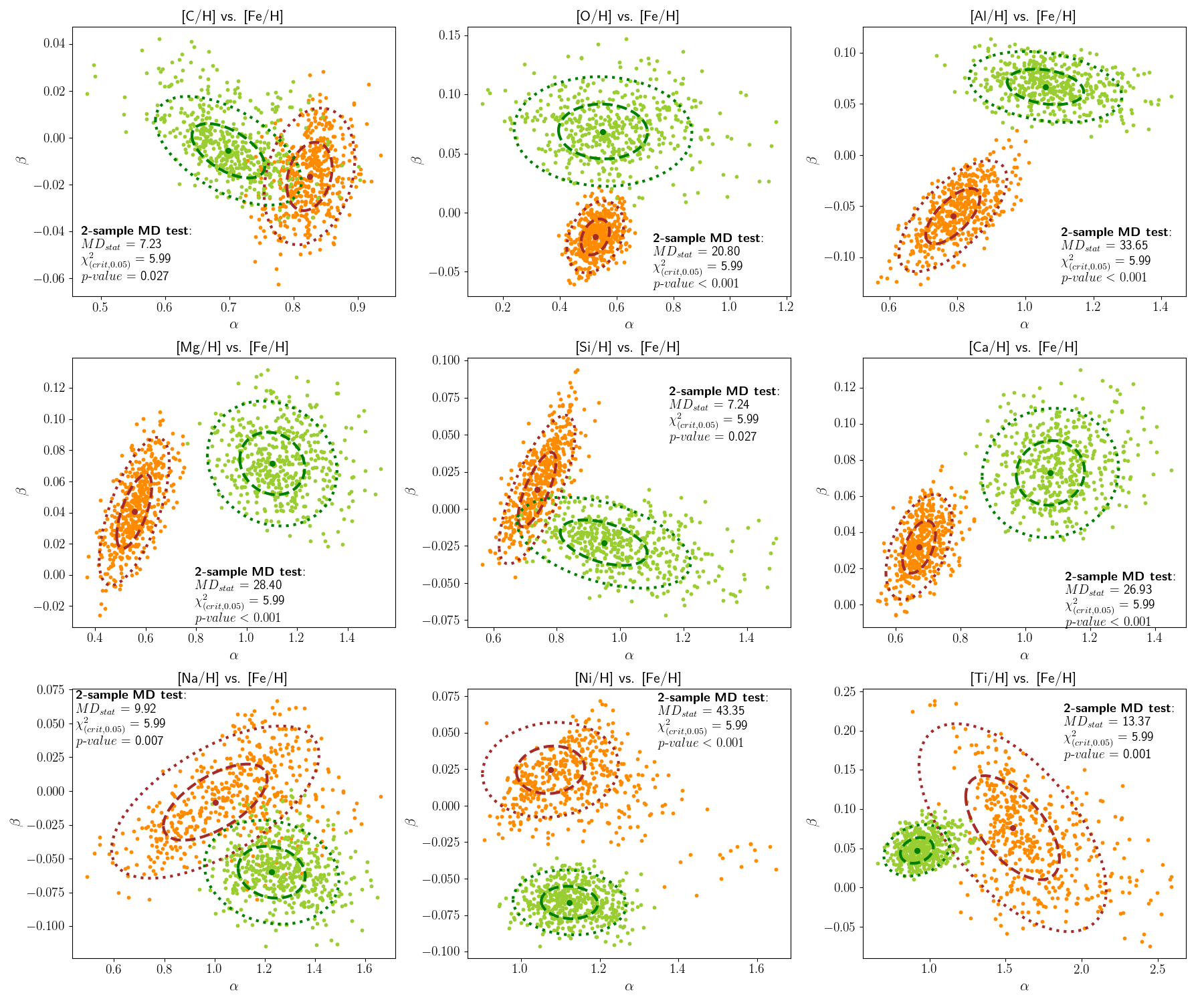}
	\caption{\edit{Similar to Fig. \ref{fig:GK_Mboot_coefs_comp_ALL}, but for the "nearby" samples of both GK dwarfs (green dots) and M dwarfs (orange dots) confined to 150 pc from the Sun.}} 
	\label{fig:GKnear_Mboot_coefs_comp_ALL}
\end{figure*}

\edit{The same procedure of statistical analysis as done for the full sample of GK (thin-disc) dwarfs and the selected sample of M dwarfs has been applied for the "nearby" sample of GK (thin-disc) dwarfs and the selected sample of M dwarfs (both confined to 150 pc from the Sun). The null hypothesis tests (shown on Fig. \ref{fig:GKnear_Mboot_coefs_comp_ALL}) have reached the same statistical conclusion, i.e. "rejection/non-rejection of $H_0$", as that has been reached in the full-sample analysis for all elements, except for Na. With the "nearby" scenario, the $H_0$ for Na is also rejected.} 

\clearpage
\section{A supplementary analysis with the chemical abundances of both GK and M dwarfs from the same survey, APOGEE}\label{app:apogee}
Here, we have compiled the chemical abundances of both GK and M dwarfs from the same survey, APOGEE. The GK dwarf data come from the APOGEE DR17 \citep{Abdurrouf2022}, filtered for reliable analyses with recommended flags (\flagfont{STARFLAG=0} \& \flagfont{ANDFLAG=0}) as well as named flags (\flagfont{FE\_H}, \flagfont{X\_FE}, etc.) for individual elements. We have also only selected GK dwarfs with [Fe/H] within [-1,1] dex, as of the primary range of metallicity for our GALAH sample in the main analysis. The resultant total number of these selected APOGEE GK dwarfs is 97,136. Comparably, the M dwarf data come from the series of papers by \cite{Souto2017, Souto2018, Souto2022} for the APOGEE M-dwarf programme. As shown already in Fig. \ref{fig:M_data_scatter_werr}, the maximum number of these APOGEE M dwarfs is 24, with the number of them for Ti being only 3. Therefore, we have (1) discarded Ti for this supplementary analysis with the APOGEE data only (this also takes into account that Ti abundance is labeled `deviant' in the APOGEE DR7, which our GK dwarf sample is based on); (2) adjusted the subset size of individual resampled GK dwarfs to 25, comparable to those for the APOGEE M dwarf sample.

\begin{figure*}[htbp!]
	\centering 
	\includegraphics[trim=0cm 0cm 0cm 0cm, scale=0.4, angle=0]{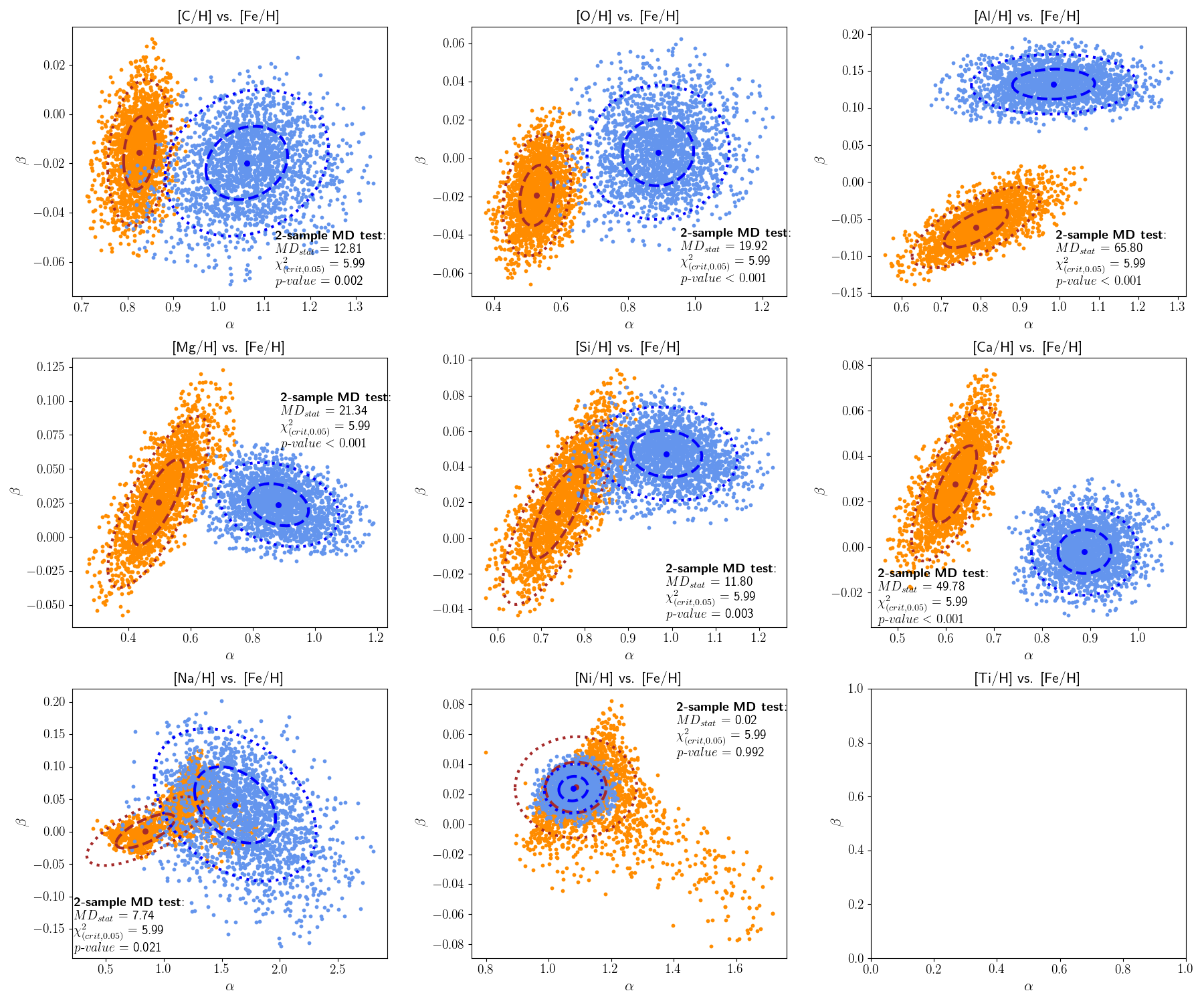}
	\caption{\edit{Similar to Fig. \ref{fig:GK_Mboot_coefs_comp_ALL}, but it is based on the samples of both GK dwarfs (blue dots) and M dwarfs (orange dots) from the same survey, APOGEE.}} 
	\label{fig:apo_GK_Mboot_coefs_comp_ALLL}
\end{figure*}

\edit{Similarly, we have performed resampling for GK dwarfs and bootstrapping for M dwarfs 2000 times, along with ODR fittings in the [Fe/H]--[X/H] space. We have also applied the same outlier identification and removal approach. The final sets of model coefficients for the selected samples of APOGEE GK and M dwarfs are compared in Fig. \ref{fig:apo_GK_Mboot_coefs_comp_ALLL}.}

\edit{The statistical test results indicate that the conclusion to reject $H_0$ reached in the main analysis remains valid for C, O, Mg, Si, Al, and Ca. However, different conclusions are reached for Na and Ni. It is noteworthy that the non-normal distributions of the model coefficients for the M-dwarf sample may affect the robustness of the test outcomes with a Mahalanobis Distance analysis, which assumes normal or quasi-normal distributions. To further examine the statistical outcomes for Na and Ni, a non-parametric test, such as a permutation test, might be useful. However, to maintain a self-consistent statistical approach across different cases, we have decided to leave this further investigation for future work.} 

\clearpage
\section{Sensitivity test for a [Mg/Si]-[C/O] trend of the M dwarf sample}\label{app:sens_test}

\begin{figure}[htbp!]
	\centering 
	\includegraphics[trim=0cm 0cm 0cm 0cm, scale=0.6, angle=0]{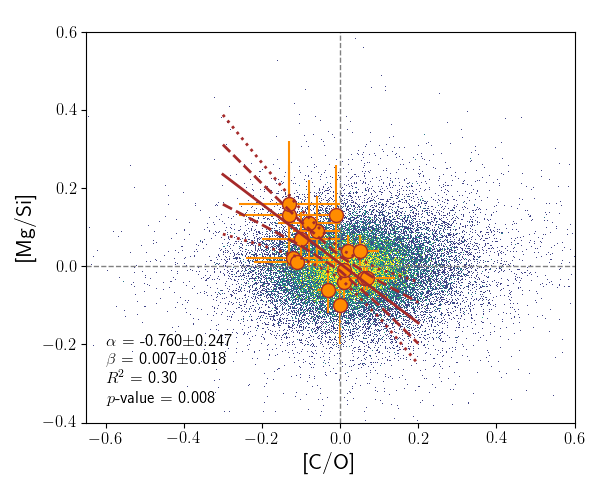} 
	\caption{\edit{Similar to Fig. \ref{fig:DR3_M_scatter_CO_MgSi}, but with those two data points of M dwarfs with [Mg/Si] > 0.2 dex removed.}}
	\label{fig:DR3_M_scatter_CO_MgSi_sens_test}
\end{figure}

\edit{We have done a sensitivity test for a [Mg/Si]-[C/O] trend of the M dwarf sample by removing those two data points with [Mg/Si] > 0.2 dex. The test result (Fig. \ref{fig:DR3_M_scatter_CO_MgSi_sens_test}) shows that the slope of the resultant trend remains statistically significant.} 

\end{document}